\documentclass[pdflatex,sn-mathphys-num]{sn-jnl}

\usepackage{graphicx}
\usepackage{multirow}
\usepackage{amsmath,amssymb,amsfonts}
\usepackage{amsthm}
\usepackage{mathrsfs}
\usepackage[title]{appendix}
\usepackage{xcolor}
\usepackage{textcomp}
\usepackage{manyfoot}
\usepackage{booktabs}
\usepackage{algorithm}
\usepackage{algorithmicx}
\usepackage{algpseudocode}
\usepackage{listings}
\usepackage{tcolorbox}
\usepackage{enumitem}
\usepackage[caption=false]{subfig}
\usepackage{soul}
\usepackage{tikz}
\usetikzlibrary{arrows.meta}
\usepackage[section]{placeins}

\tcbuselibrary{listings, skins}

\soulregister\cite7
\soulregister\ref7
\soulregister\textsc7
\soulregister\textit7
\soulregister\textbf7
\soulregister\emph7
\soulregister\textsf7
\soulregister\texttt7

\lstdefinestyle{pybox}{
  language=Python,
  basicstyle=\ttfamily\scriptsize,
  keywordstyle=\bfseries\color{blue!70!black},
  stringstyle=\color{orange!80!black},
  commentstyle=\itshape\color{gray!70!black},
  numbers=none,
  xleftmargin=2pt,
  framexleftmargin=2pt,
  breaklines=true,
  showstringspaces=false,
  tabsize=4,
  columns=fullflexible,
  keepspaces=true,
}

\newcommand{\swap}[1]{\colorbox{blue!12}{#1}}
\newcommand{\fseg}{\textcolor{orange!80!black}{\rule[-0.15ex]{2.4mm}{2.4mm}}\hspace{1.2pt}}
\newcommand{\eseg}{\textcolor{black!16}{\rule[-0.15ex]{2.4mm}{2.4mm}}\hspace{1.2pt}}
\newcommand{\panelrule}{\par\smallskip{\color{black!15}\rule{\linewidth}{0.4pt}}\par\smallskip}

\graphicspath{{figures/}}

\theoremstyle{thmstyleone}

\theoremstyle{thmstyletwo}

\theoremstyle{thmstylethree}

\begin{document}

\title[Memorization Diagnostics for Code LLMs Should be Scale-Aware]{Memorization Diagnostics for Code LLMs Should be Scale-Aware}

\author*[1]{\fnm{Prateek Kumar} \sur{Rajput}}\email{prateek.rajput@uni.lu}
\author[1]{\fnm{Abdoul Aziz} \sur{Bonkoungou}}\email{abdoul.bonkoungou@uni.lu}
\author[1]{\fnm{Alberick Euraste} \sur{Djir\'e}}\email{euraste.djire@uni.lu}
\author[1]{\fnm{Xunzhu} \sur{Tang}}\email{xunzhu.tang@uni.lu}
\author[1]{\fnm{Yewei} \sur{Song}}\email{yewei.song@uni.lu}
\author[1]{\fnm{Iyiola Emmanuel} \sur{Olatunji}}\email{emmanuel.olatunji@uni.lu}
\author[1]{\fnm{El Hacen} \sur{Diallo}}\email{el-hacen.diallo@uni.lu}
\author[1]{\fnm{Jacques} \sur{Klein}}\email{jacques.klein@uni.lu}
\author[1]{\fnm{Tegawend\'e F.} \sur{Bissyand\'e}}\email{tegawende.bissyande@uni.lu}

\affil*[1]{\orgname{University of Luxembourg}, \orgaddress{\city{Esch-sur-Alzette}, \country{Luxembourg}}}

\abstract{The extent to which large language models for code rely on memorization over genuine understanding remains highly debated. While current literature frequently reports widespread memorization, evaluating the underlying probing techniques across dense architectures reveals a severe breakdown in their utility at scale. Traditional encoder-style probes using perturbations such as synonym fuzzing or dead-code insertion struggle to expose memorization in scaled models, even on known-contaminated benchmarks, and decoder-style probes that rely on log probabilities show similar performance degradation. The specific mode of failure for these probes, particularly why such techniques disrupt smaller models but fail to impact larger ones, motivates us to untangle representation load from memorization rather than treating them as a single phenomenon. By applying invertible mathematical transforms to numeric problems, we isolate these two factors and reveal that scaled encoders successfully absorb substantial representation load while still converging on the correct family of solutions. In practical software engineering, this ability to adapt to varying surface forms is what truly matters for usability and generalizability in LLM and agentic applications. Whether a specific solution was seen during training becomes a much less pressing question because although memorization inflates scores on contaminated benchmarks, factoring out representation load makes it debatable how much we should truly care if a functional answer was originally memorized. Future evaluations must therefore be built around separating these phenomena rather than relying on methodologies that quietly entangle them. Code and Data are available\footnote{\url{https://github.com/pkrajput/isomorphic_prompts}, \url{https://doi.org/10.5281/zenodo.19248561}}.}


\maketitle

\section{Introduction}\label{sec:intro}

LLMs are rapidly becoming the default engine for automating software engineering tasks, and code synthesis sits at the frontier of this transformation. They achieve impressive benchmark accuracy~\cite{chen2021evaluating, austin2021program, jimenez2023swe, zhang2025swe}, yet a persistent question shadows their progress. \emph{Are they solving problems or recognizing them?} Recent work sharpens the worry, with analyses of agentic coding benchmarks arguing that some leading results lean on recall more than reasoning~\cite{liang2025swe}. The dominant strategy in the literature for probing this question is to apply semantics-preserving perturbations to the prompt, such as renaming variables~\cite{chakraborty2022natgen, wang2023recode}, paraphrasing docstrings~\cite{mastropaolo2023robustness, chen2024nlperturbator}, and inserting dead code~\cite{chakraborty2022natgen, wang2023recode}, then reading the resulting pass@$k$ drop as evidence bearing on memorization~\cite{djire2025memorization, euraste2026learned, riddell2024quantifying}. These \emph{encoder-side} diagnostics operate in lexical and semantic space. They are intuitive and have no doubt enhanced our understanding, but they are also just convenient. Most studies say little about the extra demands these prompt changes place on the model, rarely examine scaled models where encoder capacity may absorb the perturbation entirely, and offer no formal guarantee that the rewritten prompt preserves the original task semantics. A complementary line of work targets the \emph{decoder}. Methods such as CoDeC~\cite{zawalski2025detecting} measure how in-context examples shift token log-likelihoods to distinguish seen from unseen training data, but these probes are also often validated on small model families and do not address whether their discriminative signal survives as the model's internal representation space scales and the probability mass spreads across a vastly larger set of plausible continuations. Other such decoder-style works include permutation tests on dataset ordering~\cite{oren2023proving}, multiple-choice contamination quizzes~\cite{golchin2025data}, and token-level likelihood analyses for verbatim code regurgitation~\cite{nie2025decoding}.

 We put the assumption that \emph{the probe keeps working as models grow} under direct test, and on benchmarks with known contamination~\cite{austin2021program, zawalski2025detecting, gao2020pile, raffel2020exploring}, the encoder-side and decoder-side signals we studied faded as model capacity increased. This probe saturation pushed us toward a different question. Rather than asking whether a solution was seen during training, we ask \emph{how faithfully a model carries its algorithmic knowledge through an unfamiliar way of stating the same task}. We treat this capacity as distinct from memorization and term it \emph{representational load}, and our experiments later confirm that the two are separable. The framing borrows from metamorphic testing, where a known relation between inputs constrains the relation between outputs. To add representational load while holding the core task fixed, we leverage invertible numerical transforms. We develop both ideas below.

\medskip\noindent\textbf{Representational load.}
Representational load arises wherever the interface departs from training-like surface form. Domain-specific encodings, legacy data formats, unfamiliar API contracts, and project-specific conventions all impose interface conditions that no benchmark captures. On the encoder side, it requires the model to parse auxiliary constraints alongside the core task specification, mapping a richer, less familiar prompt to the correct underlying semantic representation. On the decoder side, it requires the model to honour those constraints token by token during autoregressive generation, interleaving novel formatting logic with the algorithmic solution it has internalized. A model that can only produce correct code under training-like surface conditions offers weaker guarantees than one that can transport its algorithmic knowledge through an unfamiliar representational channel. Building on this view, we first study one representative probe from each side of the literature, encoder and decoder, along a model-scale axis, and then propose a new perspective that leverages a property unique to software, the latent structure of code itself.

\medskip\noindent\textbf{I/O isomorphisms.}
To probe behaviour beyond surface text manipulation, and to keep the task \emph{algorithmically fixed} under a mathematical guarantee, we use \emph{I/O isomorphisms}. Methodologically this is a form of metamorphic and robustness testing~\cite{chen1998metamorphic, segura2016survey} adapted to code LLMs. We apply a bijective value transform to the test-case integers and append an explicit encode/decode contract to the prompt, wrapping the interface in a representational layer that demands explicit computation without altering the underlying algorithmic task ($f(x) = y$). The metamorphic relation between the original test oracle and its isomorphic counterpart then serves as our oracle, so any deviation on a previously passing problem is, by construction, a representational-robustness failure rather than a change in the task itself. \emph{The model cannot memorize its way through this layer.} It must decode every input, run the algorithm, and re-encode every output. For our main experiments we instantiate the transform as an affine integer isomorphism,
\[
  T_\theta(t) = at + b, \quad a \neq 0,
  \qquad
  T_\theta^{-1}(t') = \frac{t' - b}{a},
\]
and we additionally ablate over base-conversion and cubic bijection families. Here, $t$ represents the original test-case integer input, while $a$ and $b$ are the randomly selected constants (with $a \neq 0$) that define the linear transformation parameterizing the affine isomorphism.

\medskip\noindent
Our study yields three findings.
\begin{itemize}[leftmargin=15pt]
    \item Synonym substitution in prompts and token log-likelihood probes that rely on in-context examples both saturate at scale, losing their power to separate seen from unseen problems.
    \item Representational load degrades performance even in scaled models, yet this drop does not by itself reveal memorization.
    \item Opcode analysis shows that scaled models still reach the same solution families under added representational load, only less often, while smaller models drop whole families of solutions altogether.
\end{itemize}

\begin{figure}[!t]
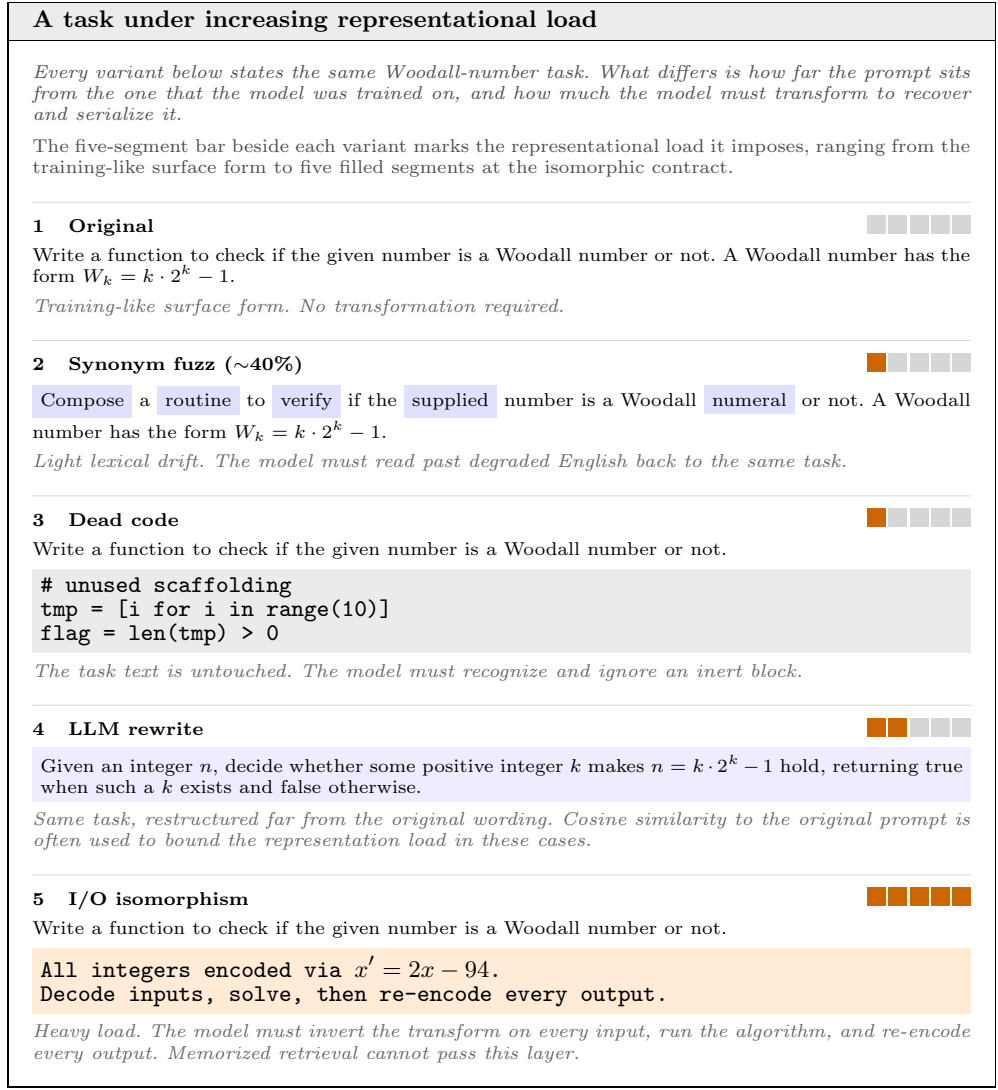

\begin{tcolorbox}[
  colback=white, colframe=black, boxrule=0.6pt,
  arc=0pt, outer arc=0pt,
  left=6pt, right=6pt, top=6pt, bottom=6pt,
  fonttitle=\small\bfseries,
  title={A task under increasing representational load},
  coltitle=black, colbacktitle=gray!15
]
\footnotesize
\noindent\textit{\color{black!70}Every variant below states the same Woodall-number task. What differs is how far the prompt sits from the one that the
model was trained on, and how much the model must transform to recover and serialize
it.}\\[4pt]
\noindent{\color{black!70}The five-segment bar beside each variant marks the representational load it imposes, ranging from the training-like surface form to five filled segments at the isomorphic contract.}
\panelrule
\noindent\textbf{1\quad Original}\hfill \eseg\eseg\eseg\eseg\eseg\\[3pt]
Write a function to check if the given number is a Woodall number or not.
A Woodall number has the form $W_k = k\cdot 2^k - 1$.\\[3pt]
{\itshape\color{black!60} Training-like surface form. No transformation required.}
\panelrule
\noindent\textbf{2\quad Synonym fuzz ($\sim$40\%)}\hfill \fseg\eseg\eseg\eseg\eseg\\[3pt]
\swap{Compose} a \swap{routine} to \swap{verify} if the \swap{supplied} number
is a Woodall \swap{numeral} or not. A Woodall number has the form
$W_k = k\cdot 2^k - 1$.\\[3pt]
{\itshape\color{black!60} Light lexical drift. The model must read past degraded
English back to the same task.}
\panelrule
\noindent\textbf{3\quad Dead code}\hfill \fseg\eseg\eseg\eseg\eseg\\[3pt]
Write a function to check if the given number is a Woodall number or not.\\[3pt]
\colorbox{black!8}{\parbox{\dimexpr\linewidth-2\fboxsep\relax}{\ttfamily\scriptsize
\# unused scaffolding\\
tmp = [i for i in range(10)]\\
flag = len(tmp) > 0}}\\[3pt]
{\itshape\color{black!60} The task text is untouched. The model must recognize
and ignore an inert block.}
\panelrule
\noindent\textbf{4\quad LLM rewrite}\hfill \fseg\fseg\eseg\eseg\eseg\\[3pt]
\colorbox{blue!7}{\parbox{\dimexpr\linewidth-2\fboxsep\relax}{Given an integer
$n$, decide whether some positive integer $k$ makes $n = k\cdot 2^k - 1$ hold,
returning true when such a $k$ exists and false otherwise.}}\\[3pt]
{\itshape\color{black!60} Same task, restructured far from the original wording.
Cosine similarity to the original prompt is often used to bound the representation load in these cases.}
\panelrule
\noindent\textbf{5\quad I/O isomorphism}\hfill \fseg\fseg\fseg\fseg\fseg\\[3pt]
Write a function to check if the given number is a Woodall number or not.\\[3pt]
\colorbox{orange!16}{\parbox{\dimexpr\linewidth-2\fboxsep\relax}{\ttfamily\scriptsize
All integers encoded via $x' = 2x - 94$.\\
Decode inputs, solve, then re-encode every output.}}\\[3pt]
{\itshape\color{black!60} Heavy load. The model must invert the transform on
every input, run the algorithm, and re-encode every output. Memorized retrieval
cannot pass this layer.}
\end{tcolorbox}
\caption{The same task posed at increasing representational load, with the five-segment bar beside each variant indicating how much load it imposes. Synonym
fuzzing, dead code, and a rewrite perturb the surface form only, while the isomorphic contract forces the model to invert a value transform, run the
algorithm, and re-encode every output.} 
\label{fig:repload_intro}
\end{figure}
Of the three, the third is the central phenomenon of this work, and Figure~\ref{fig:repload_intro} builds the intuition behind it. A single Woodall task is restated at growing distance from its training-like form, moving from light synonym and dead-code edits through a free-form rewrite to a full isomorphic contract, with each variant forcing the model to do more work to recover the task and then serialize its answer. The pass@$k$ drop that follows invites a naive reading in which the model has simply lost the algorithm, and we argue this reading is incomplete. What the added load strains is the route from a known solution to a correctly encoded one, not the model's grasp of the solution itself. Section~\ref{sec:case_woodall} makes this concrete on a single failing trace (Figure~\ref{fig:case_woodall_intro}), where the algorithmic core is identical to that of a passing run and only the output encoding breaks.

\section{Running Example}
\label{sec:case_woodall}

\begin{figure}[!t]
\begin{tcolorbox}[
  colback=white, colframe=black, boxrule=0.6pt,
  arc=0pt, outer arc=0pt,
  left=5pt, right=5pt, top=4pt, bottom=4pt,
  fonttitle=\small\bfseries,
  title={MBPP Task~18 under a full isomorphic contract ($T_\theta(t)=2t-94$)},
  coltitle=black, colbacktitle=gray!15
]
\scriptsize

\noindent\textit{\color{black!70}Failure examples from 5 generations of the same I/O contract.}

\panelrule

\noindent\textbf{Prompt and encoding contract}\\[2pt]
Write a function to check if the given number is a Woodall number or not.
A Woodall number has the form $W_k = k \cdot 2^k - 1$.\\[2pt]
\colorbox{orange!16}{\parbox{\dimexpr\linewidth-2\fboxsep\relax}{\ttfamily
All integers encoded via $x' = 2x - 94$.\\
Decode inputs as $x = (x' + 94)/2$, solve, then re-encode every output.\\
E.g.\ input $-74$ represents $10$; if the answer is $0$, output $-94$.}}

\panelrule

\noindent\textbf{Test cases:}\quad\\
\\
\begin{tabular}[t]{@{}l@{\;\;}l@{\;\;}l@{\;\;}l@{}}
\textsf{Original} & \texttt{is\_woodall(383)==True} & \texttt{is\_woodall(254)==False} & \texttt{is\_woodall(200)==False}\\
\textsf{Iso}      & \texttt{is\_woodall(672)==True} & \texttt{is\_woodall(414)==False} & \texttt{is\_woodall(306)==False}\\
\end{tabular}

\panelrule

\noindent
\begin{minipage}[t]{0.485\linewidth}
\textbf{Passing solution}\hfill{\color{black!60}3/5}\\[2pt]
\begin{lstlisting}[style=pybox]
def is_woodall(n_enc):
    n = (n_enc + 94) / 2
    if n <= 0 or not n.is_integer():
        return False
    n = int(n)
    k = 1
    while True:
        w = k * (2**k) - 1
        if w == n:  return True
        if w > n:   return False
        k += 1
\end{lstlisting}
\end{minipage}\hfill
\begin{minipage}[t]{0.485\linewidth}
\textbf{Failing solution}\hfill{\color{black!60}2/5}\\[2pt]
\begin{lstlisting}[style=pybox, escapechar=|]
def is_woodall(enc_n):
    n = (enc_n + 94) / 2
    if not n.is_integer():
        return (0*2)-94 |\textcolor{red!75!black}{== -94}|
    n = int(n)
    if n <= 0:
        return (0*2)-94 |\textcolor{red!75!black}{== -94}|
    k = 1
    while True:
        w = k * (2**k) - 1
        if w == n:
            return (1*2)-94 |\textcolor{red!75!black}{== -92}|
        if w > n:
            return (0*2)-94 |\textcolor{red!75!black}{== -94}|
        k += 1
\end{lstlisting}
\end{minipage}

\panelrule

\noindent\textit{\color{black!75}\textbf{Why it fails.} The algorithmic loop is
identical in both runs. The break is decoder-side: complying with the integer
contract token-by-token, the model routes every boolean answer through an
\emph{encoded} arithmetic expression
(\texttt{(b*2)-94\,==\,...}) which in Python evaluates back to
\texttt{True}, so all four returns collapse to the same value. The load of
\emph{following the contract during generation}, not the algorithm, is what
fails, and this is the load our RQs target.}

\end{tcolorbox}
\caption{Example: MBPP Task~18 under \textsc{Iso} contract. The failure mode is far from memorization, under heavy representational load.}
\label{fig:case_woodall_intro}
\end{figure}

Figure~\ref{fig:case_woodall_intro} traces one MBPP task under the isomorphic
contract. Across both the passing and
failing generations, the \texttt{while} loop that enumerates Woodall candidates
$k \cdot 2^k - 1$ is \emph{identical}. What separates a pass from a failure is
the return statement. The failing generation pushes its boolean answer through
the integer contract, writing comparisons such as
\texttt{(0\,*\,2)\,-\,94\,==\,-94}. In Python, that expression is \texttt{True}
on every branch, so the function reports every input as a Woodall number. The
model has not lost the algorithm. It has confused the boundary between the
encode-decode contract and Python's own type system, and it cannot reliably
serialize \texttt{False} once the native return type is boolean.

\medskip\noindent\textbf{Narrow channels.}
This is the behaviour we name a \emph{narrow channel}. The path that carries a
model's algorithmic knowledge from specification to code stays open under
representational load, but it tightens, so the same solution families still
appear, only less often. A degraded pass@$k$ then measures a constricted route
rather than a forgotten algorithm, and the Woodall trace is one instance of it,
with the correct output surviving in three of five samples, while the loop never
changes. Smaller models behave differently. Their channel closes rather than
narrows, and entire solution families drop out, which is the collapse we
separate from the graceful narrowing seen at scale.

The same trace also tells the effect apart from memorization. A retrieved
training solution should shatter under the isomorphic prompt, whose transformed
I/O does not resemble any plausible training instance, yet the scaled models
keep the task intact and break only at the contract. Prior probes would log
this is nothing more than a lower pass@$k$~\cite{euraste2026learned,
djire2025memorization, riddell2024quantifying}. To recover the algorithmic core
that pass@$k$ hides, we turn to opcode divergence between solutions under the
original and isomorphic prompts, measured over passing solutions and overall
generations.

\section{Related Work}
\label{sec:related}

\subsection{Memorization and Contamination in Code}
That memorization is real and measurable in large language models is by now well documented. Training data can be extracted near-verbatim from large models~\cite{carlini2021extracting}, and deduplicating the corpus reduces it while improving downstream behaviour~\cite{lee2022deduplicating}. For code in particular, contaminated benchmarks have been tied to inflated pass rates~\cite{riddell2024quantifying}, models score higher on problems released before their training cutoff than on fresh ones~\cite{coignion2024performance}, and the leakage of evaluation sets into training has proven both common and easy to overlook~\cite{matton2024leakage, sainz2023nlp}. The concern extends to the agentic setting, where recent analysis argues that some leading SWE benchmarks reflect recall rather than reasoning~\cite{liang2025swe}.

\subsection{Probes targeting the Encoder and Decoder}
On the encoder, the standard probe perturbs the prompt while preserving its meaning, then reads the pass@$k$ drop as a memorization signal, through variable renaming~\cite{chakraborty2022natgen, wang2023recode}, docstring paraphrase~\cite{mastropaolo2023robustness, chen2024nlperturbator}, or dead-code insertion~\cite{chakraborty2022natgen, wang2023recode}. As we argue in Section~\ref{sec:intro}, these manipulations stay in lexical and semantic space, rarely control for the extra parsing they impose, and carry no guarantee that the rewritten prompt preserves the original task. A complementary line attacks the question on the decoder. Membership inference asks whether a datum was seen, inferring it from the lowest-probability tokens of a sequence~\cite{shi2023detecting} or from how a model scores a text against close neighbours~\cite{mattern2023membership}, while counterfactual memorization~\cite{zhang2023counterfactual} and influence functions~\cite{grosse2023studying} attribute behaviour to individual training examples. These methods sharpen what can be said about whether a specific input was seen, yet they answer a membership question rather than the capability question a practitioner cares about, which we try to target via separating the representational load.

\subsection{Probe Reliability at Scale}
Memorization is known to grow with model size and data duplication~\cite{carlini2021extracting} and to emerge in partly predictable, scale-dependent ways~\cite{biderman2023emergent}, so a probe calibrated on a small family is reading a moving quantity. Membership-advantage probes make this concrete, having been validated on a handful of small, open-code models where they separate likely-seen inputs from unseen ones~\cite{euraste2026learned, djire2025memorization}, though whether that separation survives at the scales now deployed in practice remains untested. Training dynamics complicate the picture further. Grokking shows that models can fit their data long before they generalize, with generalization arriving only after extended training uncovers latent structure~\cite{power2022grokking}, an effect later traced to the gradual formation of structured internal circuits~\cite{nanda2023progress} and explained as a shift toward the most efficient representation~\cite{liu2022towards}, while memorization itself accrues across training in ways that do not track simple overfitting~\cite{tirumala2022memorization}. The duration and composition of training, not model size alone, may therefore be a decisive and largely uncontrolled variable, since two models of the same size can sit on opposite sides of a grokking-like transition.

\subsection{Moving Targets and Our Perspective}
The community's main response to contamination has been to build moving targets, whether contamination-free live benchmarks~\cite{jain2024livecodebench}, harder test suites that expose brittle pass rates~\cite{liu2023your}, or benchmarks that evolve known problems into fresh variants~\cite{xia2024top}. We take a different route. Because capability rises predictably with scale~\cite{kaplan2020scaling} and larger models adapt to unfamiliar prompt specifications through in-context learning~\cite{wei2022emergent, brown2020language}, we hold the algorithm fixed and change only how it is stated, asking what survives once representational load is separated from recall rather than chasing ever-cleaner data.

\section{Research Questions}\label{sec:rqs}

What remains is to turn this intuition into a structured investigation, and we
proceed by elimination. If the field's standard probes still separated seen
problems from unseen ones at frontier scale, the prevailing memorization reading
would hold and little would need revising. We therefore put those probes to the
test first, one on each side of the model, and treat their continued discriminative
power as the hypothesis to be disproved. Only once that reading is shown to break
do we introduce a load the model cannot have encountered in training, and ask what
the resulting failure really is, whether the model has lost the algorithm or only
the route that serializes it. The two leave the same trace in pass@$k$, so nothing
short of this ordering can tell them apart. Three research questions follow that path, two that close out the inherited account and one that opens the one we propose.

\noindent\textbf{RQ1: To what extent do encoder-side memorization probes work at scale?}
Synonym substitution and dead-code insertion stress lexical surface form. We test whether the resulting pass@$k$ drop still tracks any diagnostic signal in frontier models, or whether scaled encoders absorb the perturbation.

\noindent\textbf{RQ2: To what extent do decoder-side memorization probes work at scale?}
Likelihood-based probes such as CoDeC compare token-level scores with and without in-context examples. We test whether the seen-vs-unseen gap they rely on survives as the model scale grows.

\noindent\textbf{RQ3: To what extent can representational load break performance?}
Using I/O isomorphisms, we keep the algorithmic task provably fixed and only change the interface. If pass@$k$ still drops, the failure cannot be a memorization artifact and must come from the load itself.

\section{Methodology}\label{sec:methodology}

\begin{sidewaysfigure}
  \centering
  \includegraphics[width=0.95\textheight]{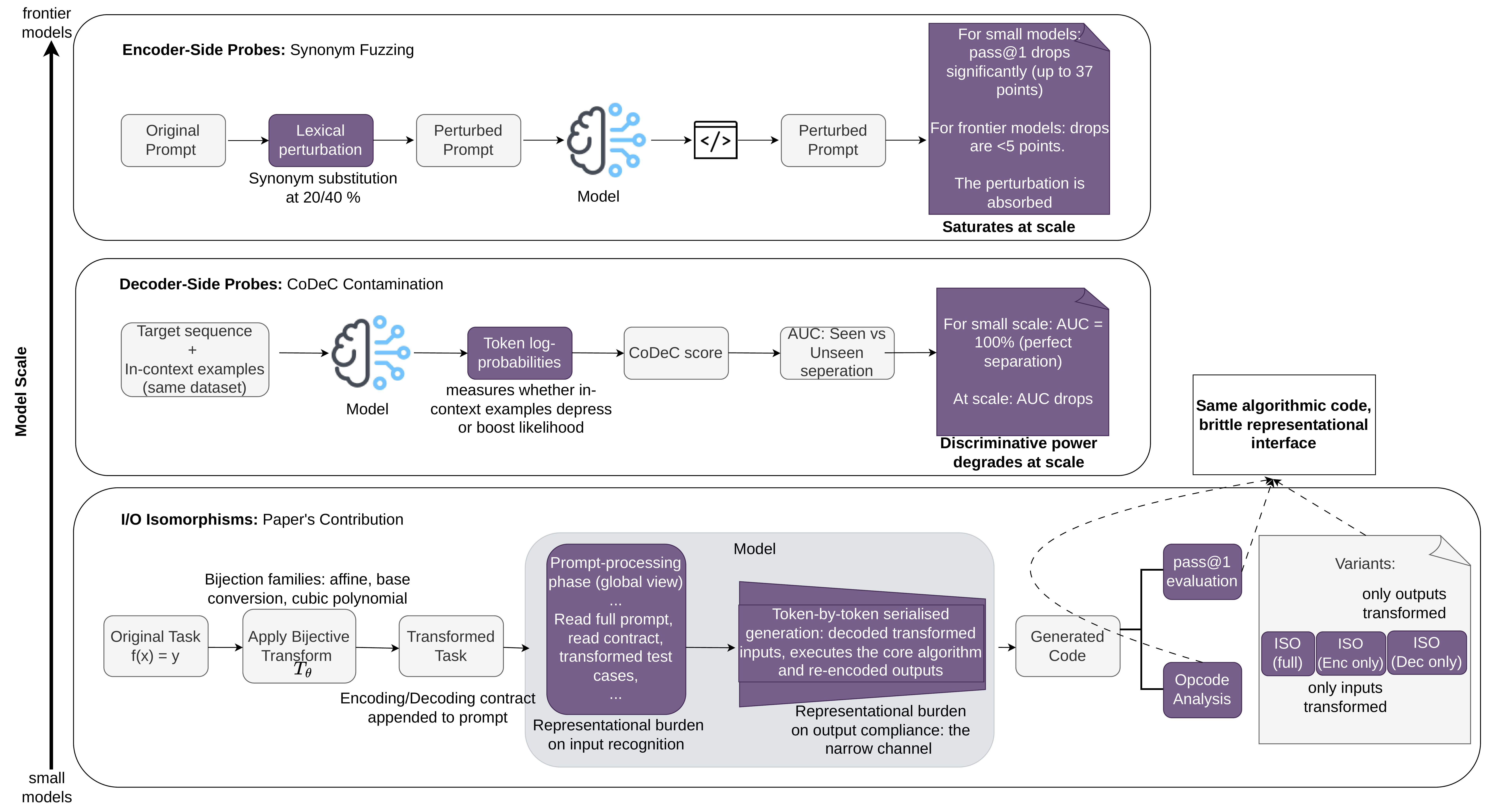}
  \caption{Overview of diagnostic protocols. We evaluate three axes across model scale, namely encoder-side synonym fuzzing (top), decoder-side CoDeC contamination probing (middle), and our proposed I/O isomorphism protocol with opcode analysis (bottom). The isomorphism preserves the algorithmic task exactly ($f(x){=}y \iff T_\theta(f(T_\theta^{-1}(x'))){=}y'$) while isolating representational load on both the prompt-processing and generation phases.}
  \label{fig:methodology}
\end{sidewaysfigure}

This section describes the experimental setup common to all three diagnostic axes, covering the models and their selection rationale (Section~\ref{sec:models}), the datasets and evaluation conditions (Section~\ref{sec:datasets}), and the protocols specific to each axis (Sections~\ref{sec:decoder_protocol}--\ref{sec:opcode_protocol}). An overview of the full methodology can be seen in Figure~\ref{fig:methodology}.

\subsection{Models}\label{sec:models}

\begin{table}[!htbp]
\footnotesize
\setlength{\tabcolsep}{5pt}
\renewcommand{\arraystretch}{1.20}
\caption{Models used in our experiments. \textbf{(a)}~Encoder-side lexical probing and I/O isomorphism experiments. \textbf{(b)}~Decoder-side experiments (CoDeC).}\label{tab:models}
\centering
\begin{minipage}[t]{0.48\linewidth}\centering
\textbf{(a)}\label{tab:models_encoder}\\[2pt]
\begin{tabular}{@{}l r r@{}}
\toprule
\textbf{Model} & \textbf{Params} & \textbf{Context} \\
\midrule
\multicolumn{3}{@{}l}{\textit{Open-weight}} \\
deepseek-coder-6.7b-instruct      & 6.7B  & 16K  \\
Llama-3.1-8B-Instruct             & 8B    & 128K \\
CodeLlama-13B-Instruct            & 13B   & 16K  \\
StarCoder2-15B                    & 15B   & 16K  \\
Codestral-22B-v0.1                & 22B   & 32K  \\
Qwen2.5-Coder-32B-Instruct        & 32B   & 128K \\
Llama-3.1-70B-Instruct            & 70B   & 128K \\
\midrule
\multicolumn{3}{@{}l}{\textit{Closed-source}} \\
gpt-4o-mini-2024-07-18            & ---   & 128K \\
gpt-4o-2024-08-06                 & ---   & 128K \\
gemini-2.0-flash                  & ---   & 1M   \\
gemini-2.5-flash                  & ---   & 1M   \\
\botrule
\end{tabular}
\end{minipage}\hfill
\begin{minipage}[t]{0.48\linewidth}\centering
\textbf{(b)}\label{tab:models_decoder}\\[2pt]
\begin{tabular}{@{}l r r@{}}
\toprule
\textbf{Model} & \textbf{Params} & \textbf{Context} \\
\midrule
\multicolumn{3}{@{}l}{\textit{Open-weight}} \\
Pythia-410M                        & 0.4B  & 2K   \\
Pythia-1.4B                        & 1.4B  & 2K   \\
Pythia-12B                         & 12B   & 2K   \\
Llama-3.1-70B                      & 70B   & 128K \\
Nemotron-4-340B-Instruct           & 340B  & 4K   \\
Llama-3.1-405B-Instruct            & 405B  & 128K \\
\midrule
\multicolumn{3}{@{}l}{\textit{Closed-source}} \\
davinci-002                        & ---   & 16K  \\
\botrule
\end{tabular}
\end{minipage}
\end{table}

We select 11 models spanning a wide range of capabilities for the encoder-side and isomorphism experiments (Table~\ref{tab:models}a), prioritizing models commonly used for coding tasks and covering a broad parameter-count range. For the decoder-side contamination analysis (Section~\ref{sec:decoder_protocol}), we construct a separate scale axis of 7 models (Table~\ref{tab:models}b) that permit \emph{teacher-forced scoring}, computing per-token log-probabilities of a target sequence conditioned on its ground-truth prefix, without generating new tokens. For open-weight models (Pythia 410M, 1.4B, 12B; Nemotron-4-340B; Llama-3.1-405B), we use the vLLM library and access prompt-token log-probabilities directly through \texttt{prompt\_logprobs}. For davinci-002, we use OpenAI's legacy completions endpoint, which exposes equivalent functionality via \texttt{echo=True} with \texttt{logprobs}, to include at least one commercial model in our study.

A deliberate design choice guides our model selection. All models in our scale axis are dense architectures, not mixture-of-experts (MoE). Because MoE models activate only a fraction of their parameters per token, the relationship between nominal parameter count and effective capacity is ambiguous. A 340B MoE model may behave more like a much smaller dense model on any given input. Since our analysis aims to study how memorization and generalization behaviours change as a function of scale, dense architectures provide a cleaner signal, as total parameters directly reflect the capacity available during inference. How decoder-side diagnostics should be adapted for MoE architectures, where routing decisions may interact with memorization patterns in ways that dense models do not exhibit, remains an open question.

\subsection{Datasets, Conditions, and Generation Setup}\label{sec:datasets}

\begin{table}[!htbp]
\centering
\footnotesize
\caption{Evaluation datasets across the study's two diagnostic axes. The upper block covers the I/O isomorphism and encoder-side perturbations. The lower block covers the CoDeC probe, split into corpora that postdate every model's training cutoff (\emph{Unseen}) and corpora that are documented or probable members of the training mixtures (\emph{Seen}).}\label{tab:datasets}
\setlength{\tabcolsep}{6pt}
\renewcommand{\arraystretch}{1.15}
\begin{tabular}{@{}llll@{}}
\toprule
Dataset & Tasks / Role & Domain & Released \\
\midrule
\multicolumn{4}{@{}l}{\textit{I/O isomorphism and encoder-side perturbations}} \\
MBPP~\cite{austin2021program}            & 257 & Mostly-basic Python    & Aug 2021 \\
EffiBench~\cite{huang2024effibench}      & 338 & Efficiency             & Nov 2024 \\
BigOBench~\cite{chambon2025bigo}         & 640 & Algorithmic complexity & Mar 2025 \\
\midrule
\multicolumn{4}{@{}l}{\textit{Decoder-side perturbations (CoDeC)}} \\
\texttt{gpqa\_diamond}~\cite{rein2024gpqa}              & Unseen           & Graduate-level Q\&A     & Nov 2023 \\
\texttt{livecodebench\_v5}~\cite{jain2024livecodebench} & Unseen           & Competitive programming & Jan 2025 \\
\texttt{wikipedia} (Pile)~\cite{gao2020pile}            & Seen             & Encyclopedic text       & Dec 2020 \\
\texttt{hackernews} (Pile)~\cite{gao2020pile}           & Seen             & Web forum text          & Dec 2020 \\
\texttt{cc\_2024\_high}~\cite{raffel2020exploring}      & Seen$^{\dagger}$ & Common-Crawl web text   & 2024 \\
\botrule
\end{tabular}

\vspace{2pt}
{\scriptsize $^{\dagger}$ Substituted for \texttt{hackernews} when probing Nemotron-4-340B-Instruct, following the corpus-overlap designation of Zawalski et al.~\cite{zawalski2025detecting}.}
\end{table}

\noindent\textbf{Datasets.} We evaluate on three Python benchmarks spanning both a difficulty axis and a contamination axis (Table~\ref{tab:datasets}). MBPP sits at the easy end of the difficulty spectrum and is among the most widely used benchmarks in the code LLM literature~\cite{jain2024livecodebench, djire2025memorization, riddell2024quantifying}. Every model in our study was released after its publication, making all potentially contaminated on this benchmark. EffiBench and BigOBench, by contrast, were published in 2024 and 2025, respectively, after the training cutoff of all models except Gemini-2.5-Flash. This creates a natural experiment. If memorization were the primary driver of performance, we would expect systematically larger \textsc{Iso} drops on the contaminated benchmark (MBPP), while drops on unseen benchmarks (EffiBench and BigOBench) should not be systematically large. We test this prediction directly in Section~\ref{sec:results}. All three benchmarks use Python exclusively, removing cross-language variation as a confound.

\noindent\textbf{Evaluation conditions.}
We utilize six conditions for inference. The first four probe representational invariance via I/O isomorphisms; the last two serve as encoder-side baselines operating in lexical space.

\noindent\textbf{Synonym fuzzing protocol.} For the \textsc{Syn-20\%} and \textsc{Syn-40\%} conditions, we replace each non-connector, non-code-token word in the natural-language portion of the prompt with a WordNet synonym at the specified probability. This preserves task semantics while altering surface form, the same axis targeted directly in recent work~\cite{djire2025memorization} and by other similar probes like variable renaming~\cite{chakraborty2022natgen, wang2023recode} and paraphrasing~\cite{mastropaolo2023robustness, chen2024nlperturbator}. We work under the same hypothesis as theirs. If a model has merely memorized a solution, altering the surface form should degrade retrieval.

\noindent\textbf{Generation setup.} For each (task, condition) pair, we generate n=5 completions at temperature T=0.0. Fixing the temperature at zero removes temperature-driven sampling as a source of variation, so that the behaviour we observe under each condition is not confounded by deliberate exploration of the output distribution. Greedy decoding is nonetheless not bitwise deterministic in practice, because floating-point reductions on accelerators are non-associative and sensitive to batch composition, and the hosted endpoints we query give no determinism guarantee, so repeated generations of the same prompt still diverge wherever two continuations lie close in probability. \emph{It is this residual nondeterminism, not temperature, that supplies the per-problem solution spread we analyse.} We generate five completions for two reasons. First, our opcode-entropy analysis (Section~\ref{sec:opcode_protocol}) reads off a per-problem distribution of solutions, and pooling opcode frequency vectors across five generations enables meaningful per-problem Jensen-Shannon divergence estimates between conditions. Second, both EffiBench and BigOBench admit multiple valid solutions to each problem, so several generations help to characterise the solution space the model explores. The same decoding regime is held fixed across the \textsc{Original} and isomorphic conditions, so any divergence we report reflects the condition rather than a change in how solutions are drawn.

\subsection{Decoder-Side Protocol (CoDeC)}\label{sec:decoder_protocol}

We reproduce CoDeC~\cite{zawalski2025detecting}, a contamination diagnostic that operates on output log-probabilities. Each dataset receives a contamination score $S_{\mathrm{CoDeC}}(D) = (1/N)\sum_i \mathbf{1}[\Delta(x_i) < 0]$, where $\Delta(x_i)$ is the change in mean token log-likelihood of target $x_i$ when adding in-context samples from the same dataset. In-context examples typically boost confidence for unseen data but may \emph{reduce} it when the dataset was memorized during training by disrupting learned retrieval patterns. The dataset-level AUC quantifies separation between seen and unseen datasets. Formally, with seen-score set $S^{+}$ and unseen-score set $S^{-}$,
\[
\mathrm{AUC} = \frac{1}{|S^{+}|\,|S^{-}|} \sum_{s^{+}\in S^{+}}\sum_{s^{-}\in S^{-}} \left[\mathbf{1}(s^{+}>s^{-}) + 0.5\,\mathbf{1}(s^{+}=s^{-})\right].
\]

To keep frontier-scale runs feasible, we use a constrained but consistent setup with two seen and two unseen datasets per run. We designate \texttt{gpqa\_diamond}~\cite{rein2024gpqa} and \texttt{livecodebench\_v5}~\cite{jain2024livecodebench} as unseen, since both postdate the training cutoff of every model in our decoding experiments. For the seen datasets, we follow the designations reported by Zawalski et al.~\cite{zawalski2025detecting} and use \texttt{hackernews} and \texttt{wikipedia} from the pile dataset~\cite{gao2020pile} (with \texttt{cc\_2024\_high}~\cite{raffel2020exploring} replacing \texttt{hackernews} for Nemotron-4-340B-Instruct due to known corpus overlap).

\subsection{I/O Isomorphism Protocol (Metamorphic Testing for Code LLMs)}\label{sec:iso_protocol}

A coding-benchmark instance defines a mapping $f:\mathcal{X}\to\mathcal{Y}$ over textual I/O representations. We define a family of invertible value transforms parameterized by $\theta$:
\[
  T_\theta:\mathcal{V}\to\mathcal{V},\qquad T_\theta^{-1}\bigl(T_\theta(v)\bigr) = v.
\]
For the main experiments, we use affine integer isomorphisms:
\[
  T_\theta(t) = at + b,\quad a\neq 0, \qquad T_\theta^{-1}(t') = \frac{t'-b}{a}.
\]
Every integer in the test I/O is mapped through $T_\theta$. Original tests $(x,y)$ become $(x',y')$ with $x'=T_\theta(x)$, $y'=T_\theta(y)$. The prompt is augmented with a \emph{contract} specifying the values of $a$ and $b$, along with instructions to decode inputs, solve, and return encoded outputs. We additionally ablate with two alternative bijection families, base conversion and cubic polynomials, to test whether the observed brittleness generalizes beyond affine transforms (described in Section~\ref{sec:ablation_iso}).

\noindent\textbf{Correctness equivalence (the metamorphic relation).} Because $T_\theta$ is bijective,
\[
  f(x)=y \;\;\Longleftrightarrow\;\; T_\theta\!\bigl(f\bigl(T_\theta^{-1}(x')\bigr)\bigr) = y'.
\]
Any accuracy difference between \textsc{Original} and Isomorphic variant (\textsc{Iso}) of a problem is therefore attributable to representational handling, not to a change in the underlying task.

\subsection{Opcode Entropy as an Algorithmic Proxy}\label{sec:opcode_protocol}

Pass@$k$ measures whether the model produces \emph{any} correct solution but says nothing about \emph{what kind}. To characterize the solution space beyond binary correctness, we compile each generated Python program to bytecode using \texttt{dis} and compute the Shannon entropy of its opcode-frequency distribution,
\[
  H(\pi) = -\sum_{o\in\mathcal{O}} \pi(o)\,\log_2\pi(o),
\]
where $\pi(o)$ is the normalized frequency of opcode $o$. Programs relying on similar algorithmic strategies produce similar opcode profiles, making entropy a lightweight proxy for the algorithmic family a solution belongs to. To quantify distributional divergence at the per-problem level, we pool the opcode frequency vectors from all 5 generations for a given (problem, condition) pair and compute the Jensen-Shannon divergence (JSD) against the corresponding pooled distribution under \textsc{Original}. JSD is bounded in $[0,1]$ (base 2) and symmetric, providing a principled measure of how far a model's solution strategy shifts under perturbation.

\section{Results}\label{sec:results}

\noindent
Across all eleven models and three benchmarks, lexical probes (Syn) and likelihood probes (CoDeC) both saturate at scale. Our metamorphic \textsc{Iso} protocol drops frontier-model pass@1 by 14--30 points even though the algorithmic task is provably unchanged, yet these are not algorithmic failures. Pooled-opcode JSD between \textsc{Original} and \textsc{Iso} generations stays near zero for scaled models, which keep the same control flow and merely mis-serialize the interface, and spikes only for smaller models on harder benchmarks, which abandon the strategy outright. 

\definecolor{pred}{RGB}{200,40,40}
\definecolor{tierF}{RGB}{100,60,180}
\definecolor{tierM}{RGB}{180,120,20}
\definecolor{tierS}{RGB}{190,50,50}

\newcommand{\F}{\textsuperscript{\scriptsize\textcolor{tierF}{\textsf{F}}}}
\newcommand{\Md}{\textsuperscript{\scriptsize\textcolor{tierM}{\textsf{M}}}}
\newcommand{\Sm}{\textsuperscript{\scriptsize\textcolor{tierS}{\textsf{S}}}}

\begin{sidewaystable}[!htbp]
\centering
\footnotesize
\caption{Pass@1 (\%) across all conditions and models. Values in parentheses show the relative change from \textsc{Original}. ${\pm}$ denotes the half-width of a 95\,\% bootstrap CI (10000 resamples over problems). Tier labels: \textcolor{tierF}{\textsf{F}}\,=\,frontier, \textcolor{tierM}{\textsf{M}}\,=\,mid-tier, \textcolor{tierS}{\textsf{S}}\,=\,small.}\label{tab:passk}
\setlength{\tabcolsep}{4pt}
\begin{tabular}{@{}ll rrrr rr@{}}
\toprule
Dataset & Model & Orig & Iso & Iso (Enc) & Iso (Dec) & Syn-20 & Syn-40 \\
\midrule
\multirow{11}{*}{BigO}
  & gemini-2.5-flash\F            & 43.2{\scriptsize$\pm$2.2} & 25.5{\scriptsize$\pm$4.1} ({-}41\%) & 35.4 ({-}18\%) & 30.1 ({-}30\%) & 45.1 (+4\%)  & 46.2 (+7\%)  \\
  & gemini-2.0-flash\F            & 65.9{\scriptsize$\pm$2.0} & 50.3{\scriptsize$\pm$4.9} ({-}24\%) & 55.4 ({-}16\%) & 54.8 ({-}17\%) & 62.2 ({-}6\%)  & 60.5 ({-}8\%)  \\
  & gpt-4o-2024-08-06\F           & 50.9{\scriptsize$\pm$2.1} & 37.2{\scriptsize$\pm$3.1} ({-}27\%) & 46.5 ({-}9\%)  & 40.1 ({-}21\%) & 49.4 ({-}3\%)  & 48.0 ({-}6\%)  \\
  & gpt-4o-mini-2024-07-18\Md     & 41.5{\scriptsize$\pm$2.0} & 24.9{\scriptsize$\pm$4.4} ({-}40\%) & 33.8 ({-}19\%) & 27.5 ({-}34\%) & 37.4 ({-}10\%) & 33.2 ({-}20\%) \\
  & Llama-3.1-70B-Instruct\Md     & 42.6{\scriptsize$\pm$2.0} & 20.9{\scriptsize$\pm$4.7} ({-}51\%) & 32.4 ({-}24\%) & 27.3 ({-}36\%) & 37.5 ({-}12\%) & 31.1 ({-}27\%) \\
  & Qwen2.5-Coder-32B-Instruct\Md & 30.1{\scriptsize$\pm$1.9} & 18.7{\scriptsize$\pm$2.8} ({-}38\%) & 24.4 ({-}19\%) & 22.3 ({-}26\%) & 27.7 ({-}8\%)  & 26.8 ({-}11\%) \\
  & Codestral-22B-v0.1\Sm         & 22.0{\scriptsize$\pm$1.7} & 12.6{\scriptsize$\pm$3.9} ({-}43\%) & 17.8 ({-}19\%) & 13.9 ({-}37\%) & 17.2 ({-}22\%) & 16.1 ({-}27\%) \\
  & StarCoder2-15B\Sm             &  0.6{\scriptsize$\pm$0.4} &  0.0{\scriptsize$\pm$0.6} ({-}100\%) &  0.3 ({-}50\%)  &  0.0 ({-}100\%) &  0.3 ({-}50\%)  &  0.0 ({-}100\%) \\
  & CodeLlama-13B-Instruct\Sm     &  2.1{\scriptsize$\pm$0.6} &  1.7{\scriptsize$\pm$1.5} ({-}19\%)  &  1.9 ({-}10\%)  &  1.8 ({-}14\%)  &  1.2 ({-}43\%)  &  0.8 ({-}62\%)  \\
  & Llama-3.1-8B-Instruct\Sm      & 11.2{\scriptsize$\pm$1.3} &  3.8{\scriptsize$\pm$2.9} ({-}66\%)  &  7.4 ({-}34\%)  &  4.7 ({-}58\%)  &  8.9 ({-}21\%)  &  2.5 ({-}78\%)  \\
  & deepseek-coder-6.7b-instruct\Sm & 14.5{\scriptsize$\pm$1.5} & 5.1{\scriptsize$\pm$3.3} ({-}65\%) & 10.2 ({-}30\%) &  6.5 ({-}55\%)  &  8.3 ({-}43\%)  &  4.8 ({-}67\%)  \\
\midrule
\multirow{11}{*}{Effi}
  & gemini-2.5-flash\F            & 26.2{\scriptsize$\pm$3.1} & 11.7{\scriptsize$\pm$5.1} ({-}55\%) & 21.0 ({-}20\%) & 15.4 ({-}41\%) & 27.0 (+3\%)  & 28.1 (+7\%)  \\
  & gemini-2.0-flash\F            & 54.4{\scriptsize$\pm$3.1} & 29.3{\scriptsize$\pm$5.8} ({-}46\%) & 45.2 ({-}17\%) & 35.0 ({-}36\%) & 53.2 ({-}2\%)  & 51.8 ({-}5\%)  \\
  & gpt-4o-2024-08-06\F           & 62.2{\scriptsize$\pm$3.3} & 33.8{\scriptsize$\pm$6.2} ({-}46\%) & 52.1 ({-}16\%) & 40.2 ({-}35\%) & 60.5 ({-}3\%)  & 58.4 ({-}6\%)  \\
  & gpt-4o-mini-2024-07-18\Md     & 48.0{\scriptsize$\pm$3.6} & 22.6{\scriptsize$\pm$5.3} ({-}53\%) & 36.5 ({-}24\%) & 26.4 ({-}45\%) & 43.7 ({-}9\%)  & 39.4 ({-}18\%) \\
  & Llama-3.1-70B-Instruct\Md     & 36.2{\scriptsize$\pm$3.2} & 13.8{\scriptsize$\pm$5.0} ({-}62\%) & 23.5 ({-}35\%) & 17.7 ({-}51\%) & 32.2 ({-}11\%) & 24.6 ({-}32\%) \\
  & Qwen2.5-Coder-32B-Instruct\Md & 39.5{\scriptsize$\pm$3.3} & 20.1{\scriptsize$\pm$4.0} ({-}49\%) & 30.8 ({-}22\%) & 27.3 ({-}31\%) & 35.9 ({-}9\%)  & 28.4 ({-}28\%) \\
  & Codestral-22B-v0.1\Sm         & 21.3{\scriptsize$\pm$2.7} & 10.7{\scriptsize$\pm$4.4} ({-}50\%) & 16.2 ({-}24\%) & 12.1 ({-}43\%) & 18.9 ({-}11\%) &  7.6 ({-}64\%) \\
  & StarCoder2-15B\Sm             &  3.9{\scriptsize$\pm$1.3} &  0.0{\scriptsize$\pm$3.1} ({-}100\%) &  1.2 ({-}69\%) &  0.3 ({-}92\%) &  0.3 ({-}92\%) &  0.6 ({-}85\%) \\
  & CodeLlama-13B-Instruct\Sm     &  0.4{\scriptsize$\pm$0.4} &  0.0{\scriptsize$\pm$0.9} ({-}100\%) &  0.2 ({-}50\%)  &  0.0 ({-}100\%) &  0.9 (+125\%)   &  0.0 ({-}100\%) \\
  & Llama-3.1-8B-Instruct\Sm      & 14.2{\scriptsize$\pm$2.2} &  3.7{\scriptsize$\pm$3.7} ({-}74\%)  &  8.8 ({-}38\%)  &  5.0 ({-}65\%)  & 11.4 ({-}20\%)  &  9.1 ({-}36\%)  \\
  & deepseek-coder-6.7b-instruct\Sm &  9.6{\scriptsize$\pm$2.0} &  2.9{\scriptsize$\pm$3.1} ({-}70\%) &  6.4 ({-}33\%) &  3.8 ({-}60\%)  &  7.8 ({-}19\%)  &  6.2 ({-}35\%)  \\
\midrule
\multirow{11}{*}{MBPP}
  & gemini-2.5-flash\F            & 77.4{\scriptsize$\pm$2.0} & 47.2{\scriptsize$\pm$3.2} ({-}39\%) & 66.6 ({-}14\%) & 54.8 ({-}29\%) & 75.3 ({-}3\%)  & 76.1 ({-}2\%)  \\
  & gemini-2.0-flash\F            & 80.8{\scriptsize$\pm$1.8} & 58.1{\scriptsize$\pm$4.1} ({-}28\%) & 71.1 ({-}12\%) & 59.0 ({-}27\%) & 79.2 ({-}2\%)  & 76.4 ({-}5\%)  \\
  & gpt-4o-2024-08-06\F           & 84.2{\scriptsize$\pm$1.8} & 58.5{\scriptsize$\pm$3.7} ({-}31\%) & 76.8 ({-}9\%)  & 62.3 ({-}26\%) & 79.1 ({-}6\%)  & 76.6 ({-}9\%)  \\
  & gpt-4o-mini-2024-07-18\Md     & 73.0{\scriptsize$\pm$2.1} & 43.8{\scriptsize$\pm$3.2} ({-}40\%) & 61.3 ({-}16\%) & 48.9 ({-}33\%) & 68.3 ({-}6\%)  & 62.0 ({-}15\%) \\
  & Llama-3.1-70B-Instruct\Md     & 71.8{\scriptsize$\pm$2.1} & 41.4{\scriptsize$\pm$3.0} ({-}42\%) & 59.6 ({-}17\%) & 47.0 ({-}35\%) & 63.9 ({-}11\%) & 44.5 ({-}38\%) \\
  & Qwen2.5-Coder-32B-Instruct\Md & 84.0{\scriptsize$\pm$1.8} & 45.4{\scriptsize$\pm$2.1} ({-}46\%) & 68.0 ({-}19\%) & 51.2 ({-}39\%) & 76.4 ({-}9\%)  & 52.9 ({-}37\%) \\
  & Codestral-22B-v0.1\Sm         & 72.0{\scriptsize$\pm$2.1} & 44.1{\scriptsize$\pm$2.8} ({-}39\%) & 59.4 ({-}18\%) & 48.2 ({-}33\%) & 64.2 ({-}11\%) & 35.1 ({-}51\%) \\
  & StarCoder2-15B\Sm             & 24.3{\scriptsize$\pm$2.0} &  0.8{\scriptsize$\pm$3.1} ({-}97\%)  & 13.6 ({-}44\%) &  4.1 ({-}83\%)  &  6.5 ({-}73\%)  &  6.4 ({-}74\%)  \\
  & CodeLlama-13B-Instruct\Sm     & 12.8{\scriptsize$\pm$1.5} &  2.5{\scriptsize$\pm$2.1} ({-}80\%)  &  7.8 ({-}39\%)  &  3.8 ({-}70\%)  & 13.2 (+3\%)     &  4.3 ({-}66\%)  \\
  & Llama-3.1-8B-Instruct\Sm      & 48.2{\scriptsize$\pm$2.4} & 14.5{\scriptsize$\pm$3.1} ({-}70\%) & 32.4 ({-}33\%) & 19.8 ({-}59\%) & 41.2 ({-}15\%)  & 23.6 ({-}51\%) \\
  & deepseek-coder-6.7b-instruct\Sm & 52.1{\scriptsize$\pm$2.4} & 16.7{\scriptsize$\pm$3.6} ({-}68\%) & 36.0 ({-}31\%) & 22.4 ({-}57\%) & 44.8 ({-}14\%) & 30.2 ({-}42\%) \\
\botrule
\end{tabular}
\end{sidewaystable}

\subsection{Encoder-Side Probes Saturate at Scale}\label{sec:res_encoder}

Table~\ref{tab:passk} reports pass@1 under \textsc{Syn-20} and \textsc{Syn-40} across all eleven models and three benchmarks. The results show a clear saturation pattern. Synonym fuzzing remains diagnostic for small and some mid-tier models, but is nearly inert at the frontier.

\noindent\textbf{Frontier models absorb synonym perturbation.}
For frontier models, pass@1 drops remain small across benchmarks and conditions. Gemini-2.0-Flash loses at most 5.4 points, while Gemini-2.5-Flash loses even less, with a maximum drop of 2.1 points and occasional gains on BigOBench and EffiBench. GPT-4o shows the same overall pattern, with modest and consistent losses. Because these perturbations follow the synonym-fuzzing protocol of Djir{\'e} et al.~\cite{djire2025memorization}, the result points to failure of lexical perturbations as a tool to study memorization, which was validated on smaller models but loses discriminative power at the frontier scale, where encoders readily collapse semantically equivalent surface forms.

\noindent\textbf{Mid-tier models show thresholded degradation.}
GPT-4o-mini, Llama-3.1-70B, and Qwen2.5-Coder-32B fall in an intermediate regime. Under Syn-20, drops are modest, but Syn-40 produces larger and benchmark-dependent losses. For example, Llama-3.1-70B drops 7.9 points on MBPP under Syn-20 but 27.3 under Syn-40, while Qwen2.5-Coder-32B shows a similar jump on MBPP. This non-linearity suggests that mid-tier encoders tolerate moderate lexical noise but degrade once perturbation density crosses a threshold.

\noindent\textbf{Small models degrade sharply, but the signal conflates fragility with memorization.}
Below 22B parameters, synonym fuzzing causes large drops across all benchmarks. Codestral-22B, for instance, loses 36.9 points on MBPP under Syn-40, 13.7 on EffiBench, and 5.9 on BigOBench, with similar trends for Llama-3.1-8B and deepseek-coder-6.7b. If synonym fuzzing primarily disrupted memorized retrieval, the largest effects should appear on MBPP, the only benchmark plausibly contaminated for all models, with smaller drops on EffiBench and BigOBench, which postdate their training cutoffs. The data does not support this. Relative losses on unseen benchmarks are often comparable or larger, but this is confounded by task difficulty that was untested in previous works. Small models already have low baseline accuracy on harder tasks, so even small absolute drops become large relative declines. The diagnostic, therefore, mixes encoder fragility with memorization and cannot cleanly separate the two.

\noindent
\begin{tcolorbox}[colback=gray!5, colframe=black!60]
\textbf{RQ1.}~\emph{To what extent do encoder-side memorization probes work at scale?}\\[3pt]
\textbf{Their diagnostic value diminishes as scale increases.} Synonym fuzzing at 20\% and 40\% rates, following the protocol of Djir{\'e} et al.~\cite{djire2025memorization}, produces $<$8-point pass@1 drops for frontier models across all three benchmarks, while mid-tier models degrade moderately and small models collapse. The probe retains some signal at smaller scales but appears to lose much of its discriminative power at the frontier, at least for the models and benchmarks we examine.
\end{tcolorbox}

\subsection{CoDeC Loses Discriminative Power at Scale}\label{sec:res_decoder}

\begin{table}[!htbp]
\setlength{\tabcolsep}{4pt}
\renewcommand{\arraystretch}{1.10}
\caption{CoDeC contamination scores across model scale. \emph{Seen} and \emph{Unseen} report the mean contamination score for datasets designated as in or absent from the training corpus. AUC quantifies the separation between the two.}\label{tab:codec_large_models}
\centering
\footnotesize
\begin{tabular}{@{}l r c c c@{}}
\toprule
\textbf{Model} & \textbf{Params} & \textbf{Seen (\%)} & \textbf{Unseen (\%)} & \textbf{AUC (\%)} \\
\midrule
Pythia-410M     & 0.4B & 71.0 & 24.0 & 100.0 \\
Pythia-1.4B     & 1.4B & 73.5 & 21.5 & 100.0 \\
Pythia-12B      & 12B  & 75.0 & 17.0 & 100.0 \\
Llama-3.1-70B   & 70B  & 8.0  & 9.0  & 25.0 \\
davinci-002     & ---  & 43.5 & 31.0 & 75.0 \\
Nemotron-4-340B & 340B & 26.0 & 22.5 & 75.0 \\
Llama-3.1-405B  & 405B & 7.5  & 2.0  & 75.0 \\
\botrule
\end{tabular}
\end{table}

\begin{figure}[!htbp]
  \centering
  \subfloat[]{\includegraphics[width=0.7\columnwidth]{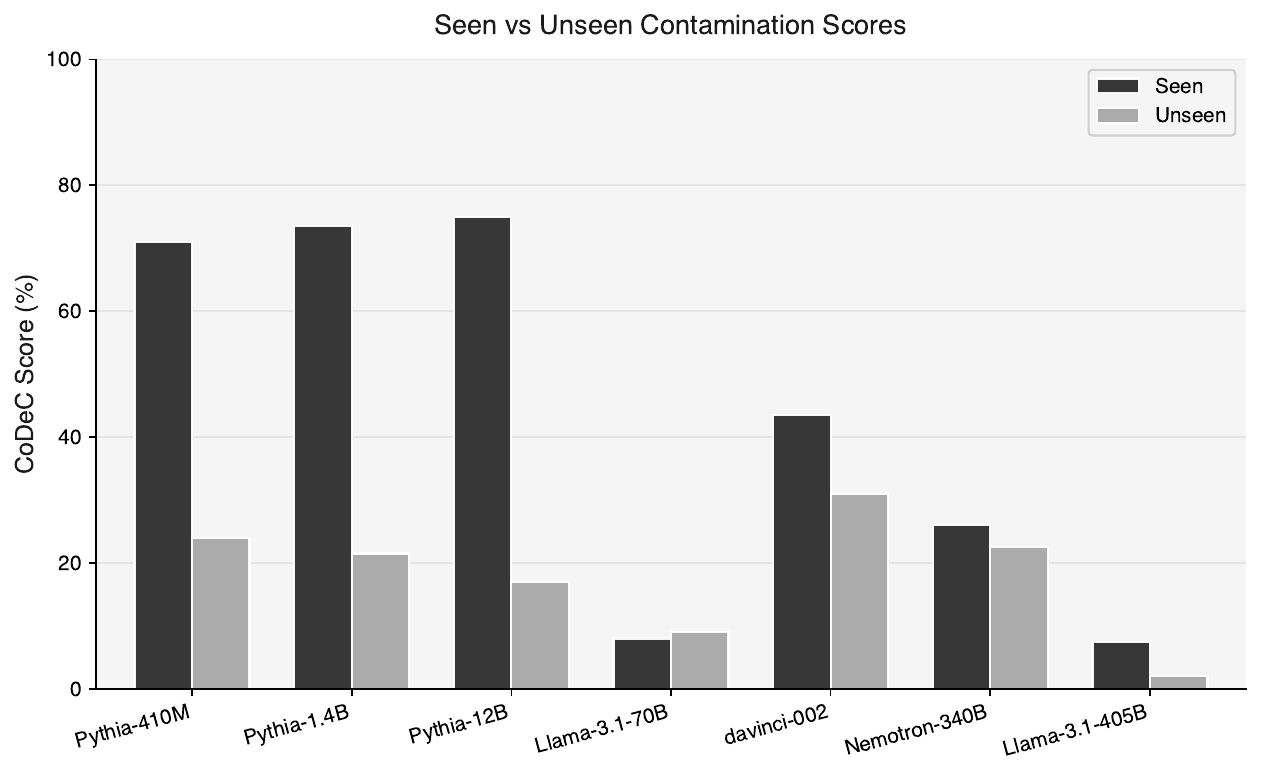}\label{fig:codec_scores}}
  \par\vspace{0.2em}
  \subfloat[]{\includegraphics[width=0.7\columnwidth]{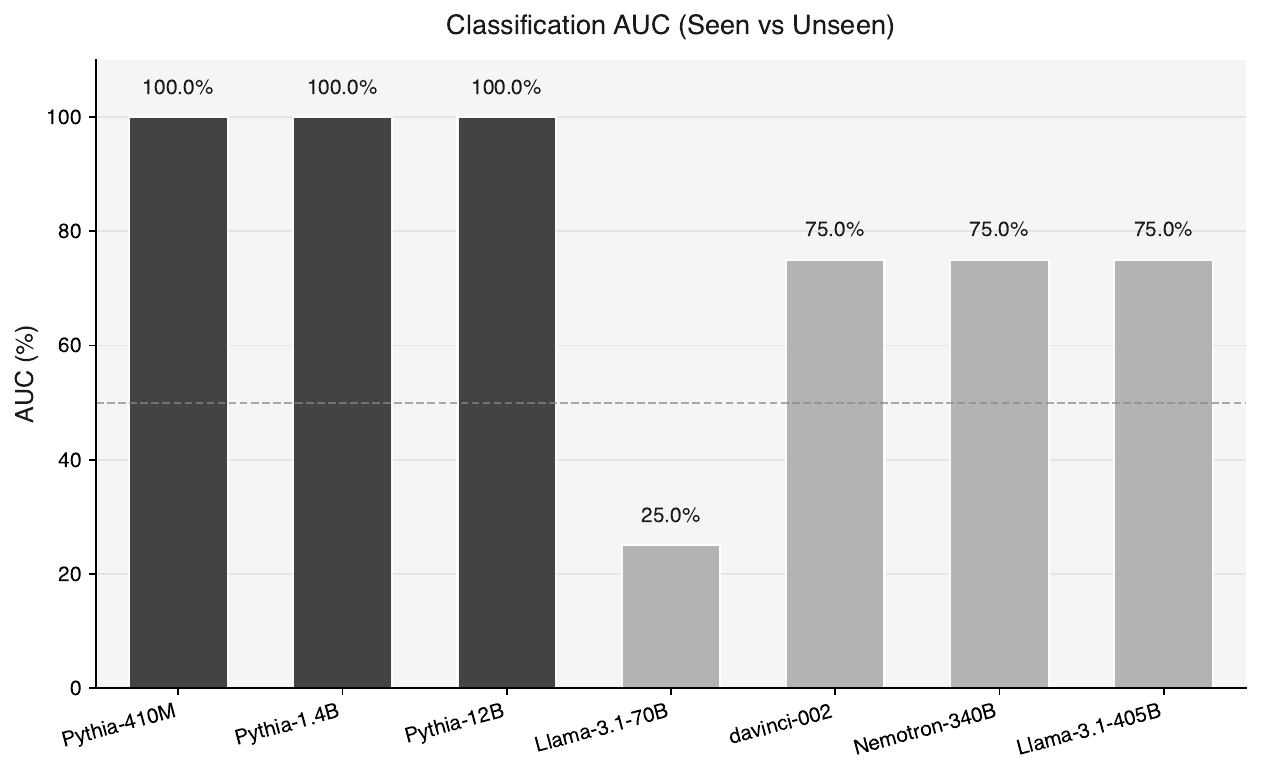}\label{fig:codec_auc}}
  \caption{(a)~Mean CoDeC scores across 4 datasets designated as \emph{Seen} or \emph{Unseen} (Section~\ref{sec:decoder_protocol}). (b)~AUC quantifying the separation between seen and unseen scores across 4 datasets.}
  \label{fig:codec}
\end{figure}

Table~\ref{tab:codec_large_models} and Figure~\ref{fig:codec} show CoDeC contamination scores across model scale. Our \emph{unseen} labels (\texttt{gpqa\_diamond}, \texttt{livecodebench\_v5}) are guaranteed, since both datasets postdate the training cutoffs of all models in our decoder-side experiments. The \emph{seen} labels (\texttt{hackernews}, \texttt{wikipedia}, and \texttt{cc\_2024\_high} for Nemotron) follow prior corpus-overlap analyses and documented training mixtures~\cite{zawalski2025detecting, gao2020pile, biderman2023pythia}. Although the seen status is therefore probable rather than certain, this asymmetry is shared across all models and does not affect the trend. Our reproduction matches the original CoDeC behaviour on Pythia: seen scores lie between 71--75\%, unseen scores between 17--24\%, and AUC remains 100\% across all three checkpoints.

\noindent\textbf{The gap reduces with scale.}
As model size increases, both contamination scores and seen-unseen separation shrink (Figure~\ref{fig:codec_scores}). davinci-002 narrows the gap to 12.5 points, Nemotron-4-340B to 3.5, and Llama-3.1-405B to 5.5. Llama-3.1-70B even inverts the expected ordering, yielding an AUC of 25\%, though this is likely an anomaly given the small number of datasets. Even ignoring that inversion, seen-dataset scores drop from 71--75\% at the Pythia scale to 7.5\% at 405B, showing that in-context augmentation shifts token log-likelihoods much less in larger models.

\noindent\textbf{Why does the signal degrade?}
CoDeC measures how in-context examples shift token log-likelihoods. Memorized targets should be disrupted, while novel targets should benefit. This assumes memorized content occupies a concentrated region of probability space that context can perturb. As models scale, that signal weakens for two reasons. First, probability mass is spread across a richer representation space, diluting concentrated retrieval effects. Second, larger models have stronger in-context learning abilities~\cite{brown2020language, wei2022emergent}, allowing them to use added context constructively for both seen and unseen targets, which reduces the asymmetry CoDeC relies on.

\noindent
\begin{tcolorbox}[colback=gray!5, colframe=black!60]
\textbf{RQ2.}~\emph{To what extent do decoder-side memorization probes work at scale?}\\[3pt]
\textbf{Their discriminative power appears to weaken with scale.} CoDeC's seen-unseen gap shrinks from 58 points (Pythia-12B) to 5.5 points (Llama-3.1-405B), and AUC drops from 100\% to 75\%. This trend is consistent with the dilution of localized retrieval signals in richer representation spaces and with the strengthening of in-context learning abilities that may override the memorization-specific asymmetry on which CoDeC depends.
\end{tcolorbox}

\begin{figure}[p]
\centering
\subfloat[Gemini-2.0-Flash:BigOBench]{\includegraphics[width=0.495\textwidth]{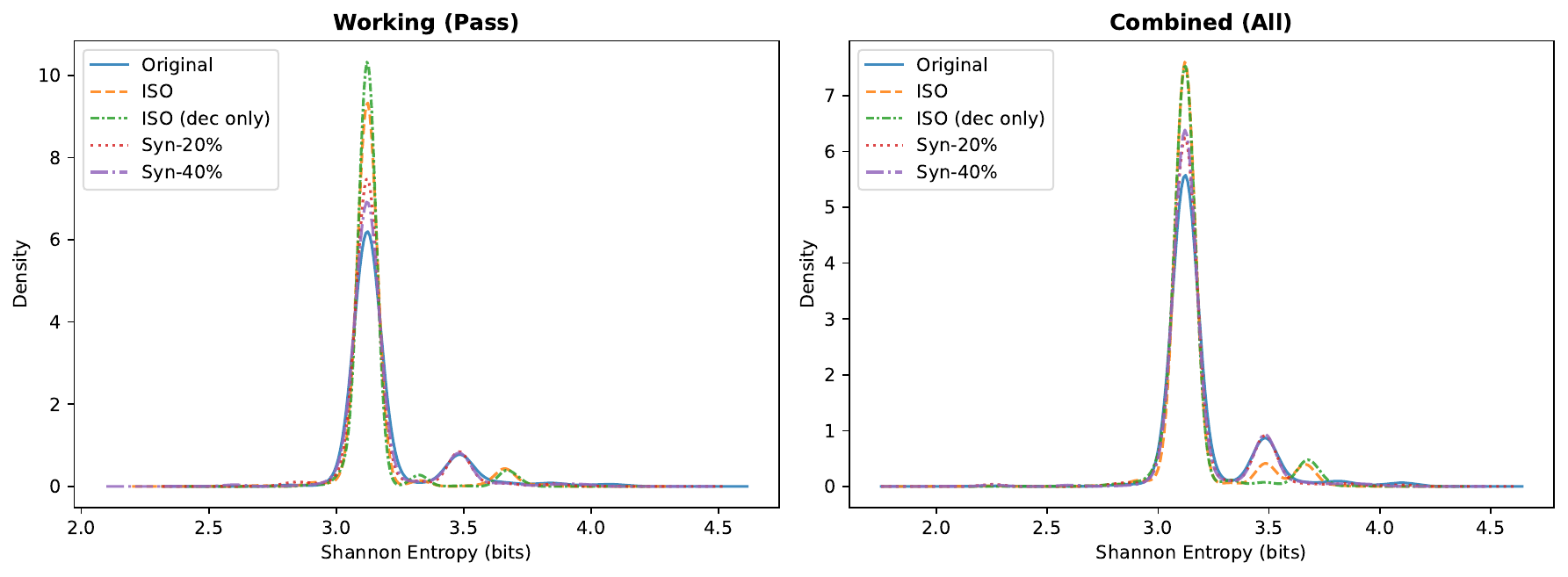}\label{fig:ent_gemini20_bigo}}\hspace{2pt}%
\subfloat[Codestral-22B:BigOBench]{\includegraphics[width=0.495\textwidth]{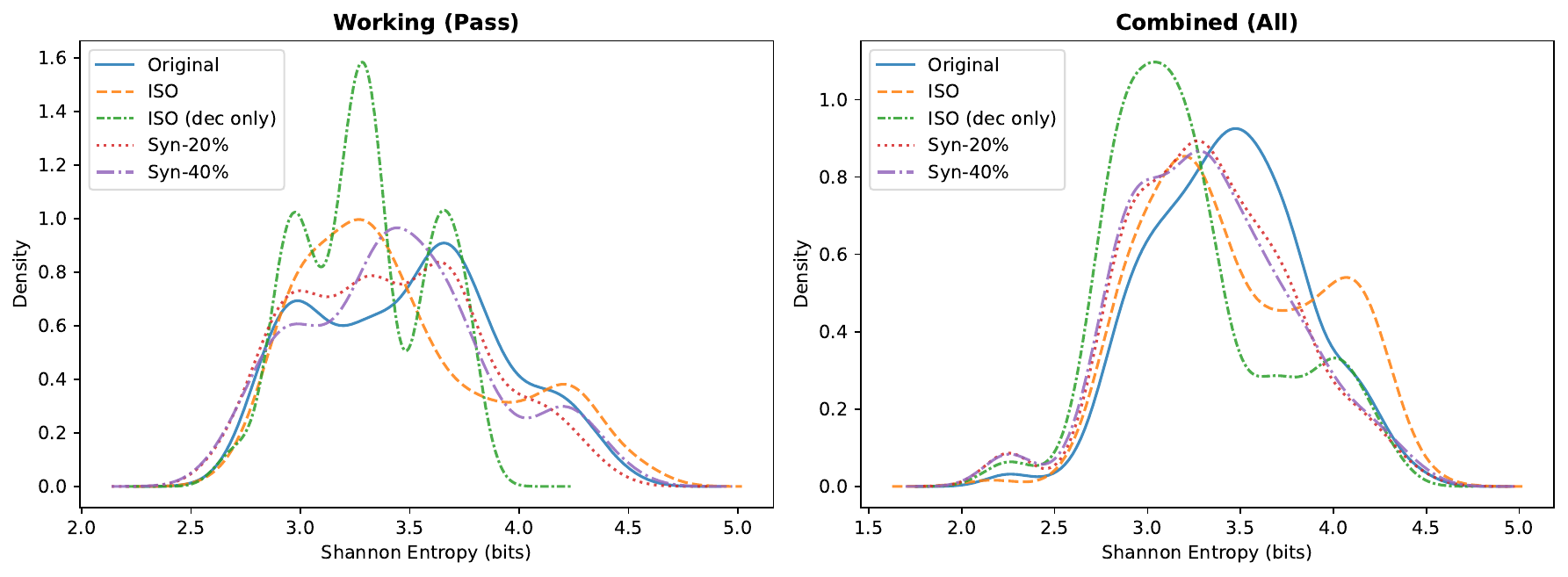}\label{fig:ent_codestral_bigo}}\\[2pt]
\subfloat[Gemini-2.0-Flash:Effibench]{\includegraphics[width=0.495\textwidth]{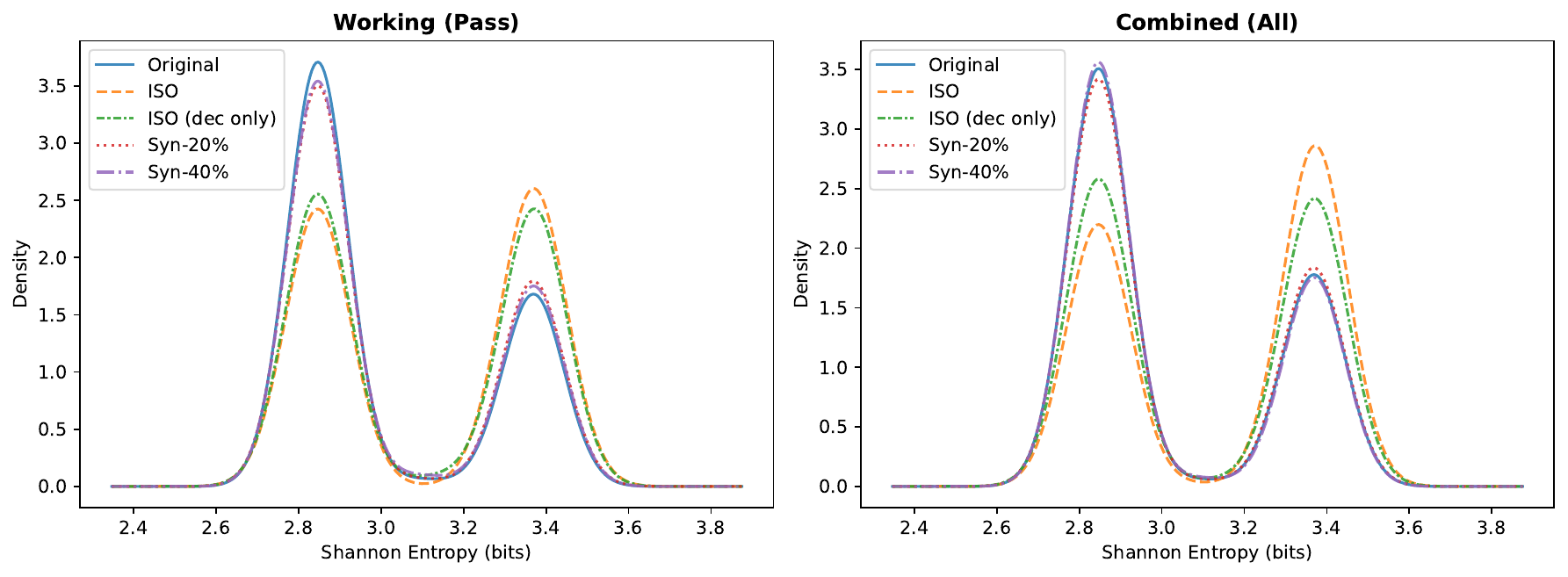}\label{fig:ent_gemini20_eff}}\hspace{2pt}%
\subfloat[Codestral-22B:Effibench]{\includegraphics[width=0.495\textwidth]{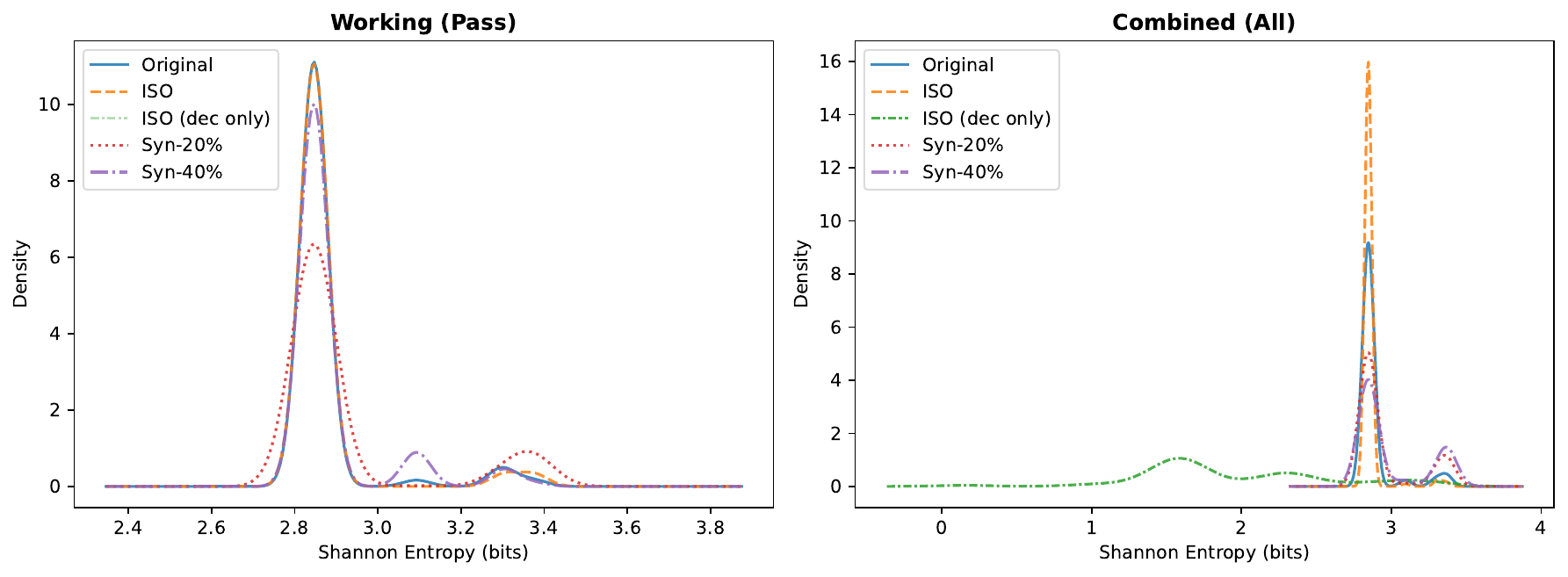}\label{fig:ent_codestral_eff}}\\[2pt]
\subfloat[Gemini-2.0-Flash:MBPP]{\includegraphics[width=0.495\textwidth]{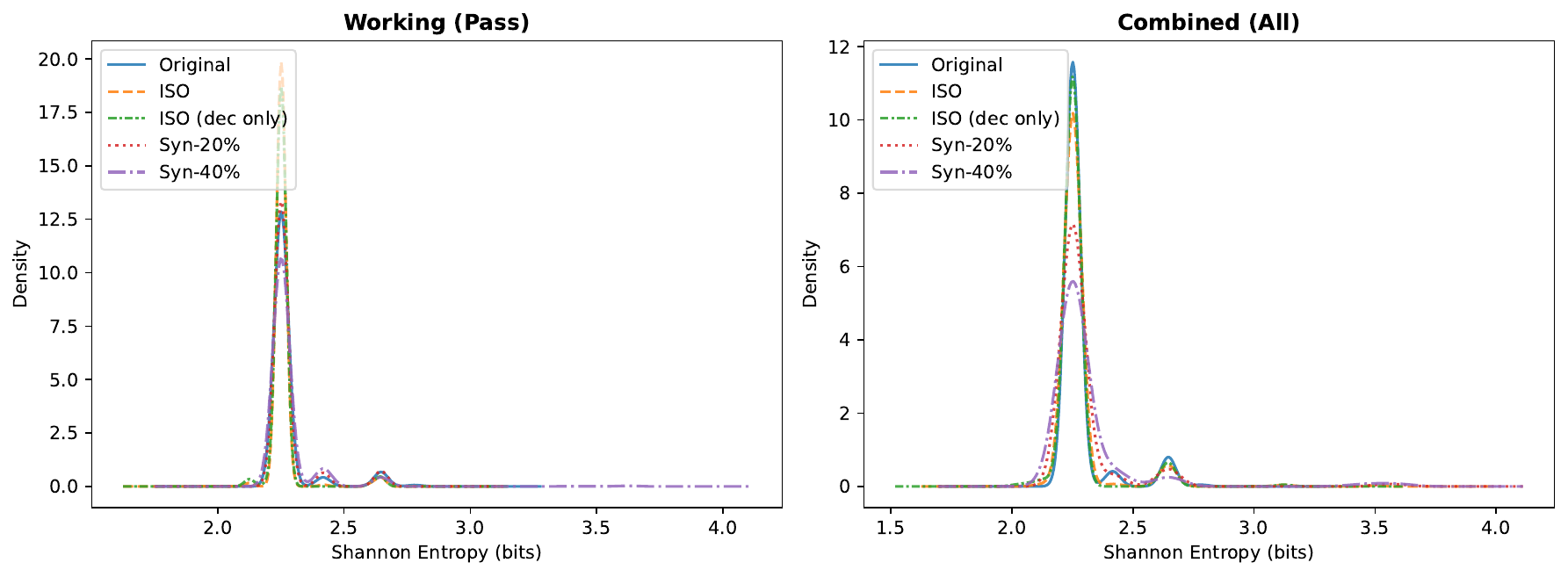}\label{fig:ent_gemini20_mbpp}}\hspace{2pt}%
\subfloat[Codestral-22B:MBPP]{\includegraphics[width=0.495\textwidth]{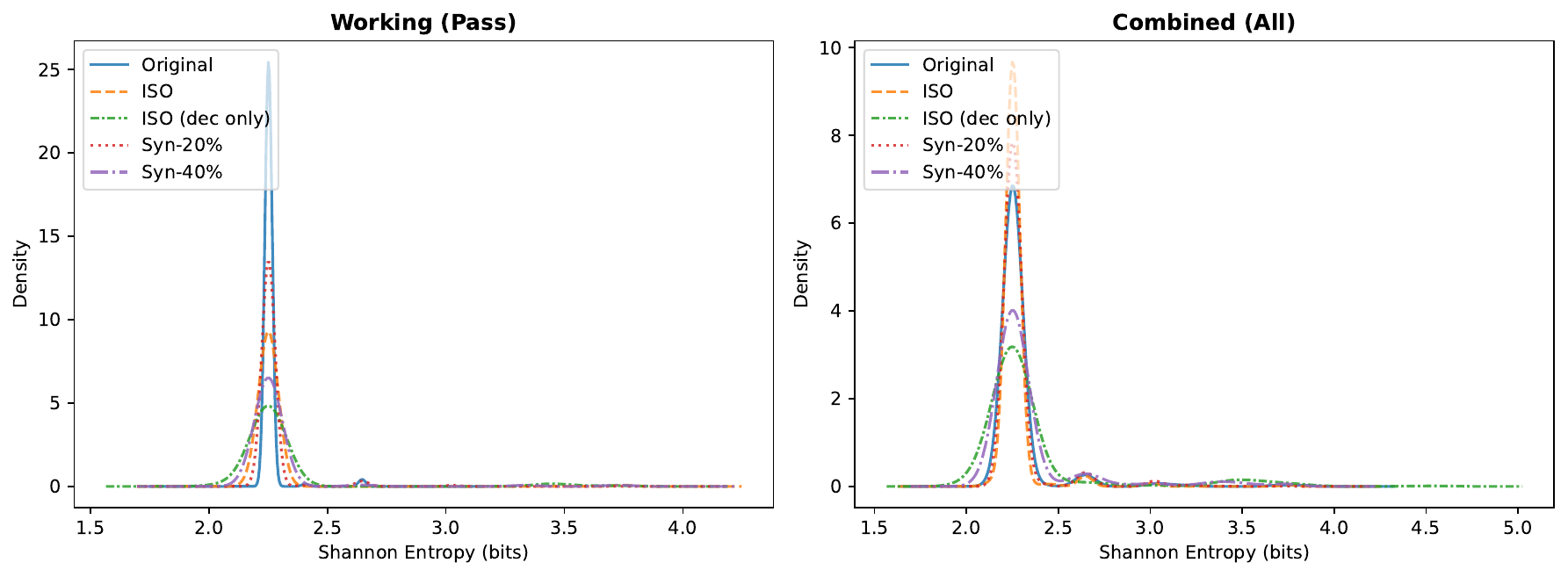}\label{fig:ent_codestral_mbpp}}
\caption{Static opcode-entropy distributions (KDE) for selected model-dataset pairs. The left column shows Gemini-2.0-flash producing nearly identical entropy profiles under \textsc{Original} and \textsc{Iso}, both for working solutions and for all generated code. The right column shows that Codestral-22b diverges substantially or produces no working solutions under \textsc{Iso} on harder datasets.}\label{fig:entropy}
\end{figure}

\begin{figure}[!htbp]
\centering
\subfloat[BigOBench]{\includegraphics[width=0.9\columnwidth]{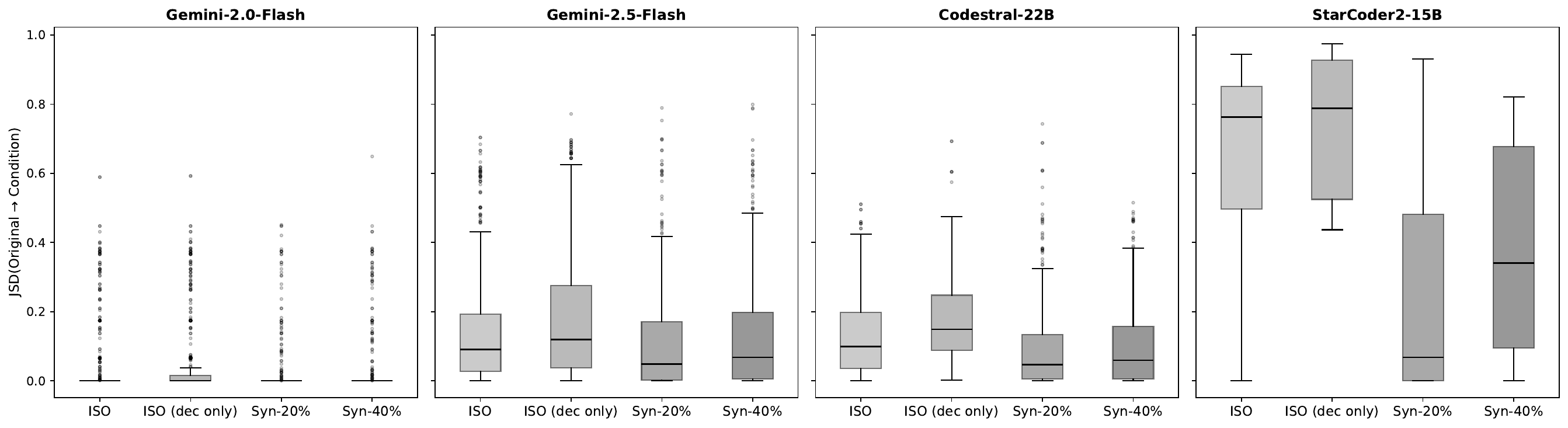}\label{fig:jsd_bigo}}
\par\vspace{0.4em}
\subfloat[EffiBench]{\includegraphics[width=0.9\columnwidth]{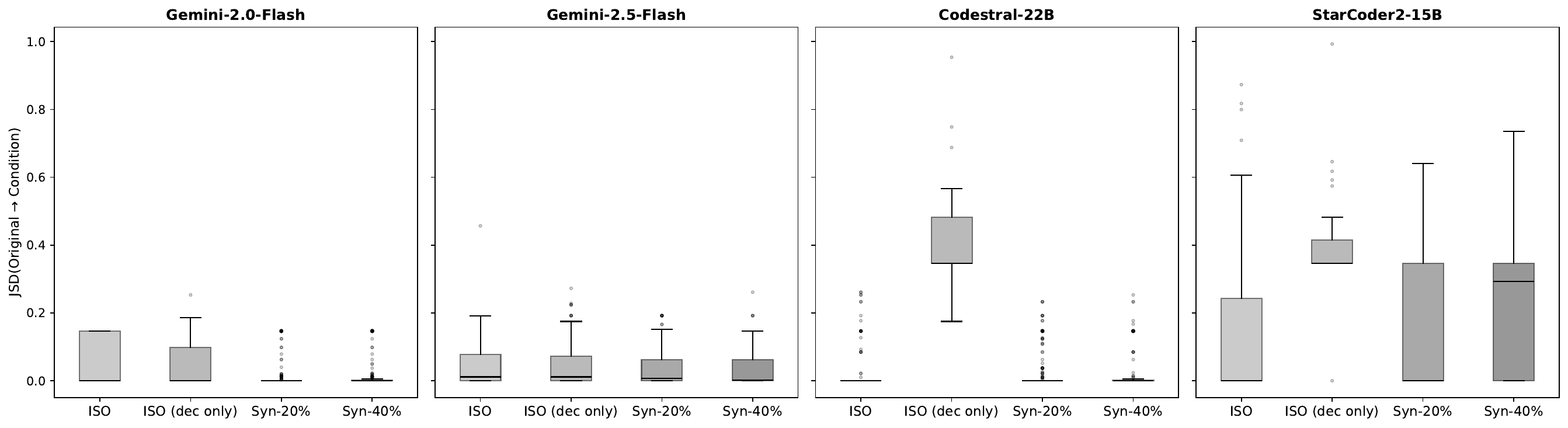}\label{fig:jsd_effi}}
\par\vspace{0.4em}
\subfloat[MBPP]{\includegraphics[width=0.9\columnwidth]{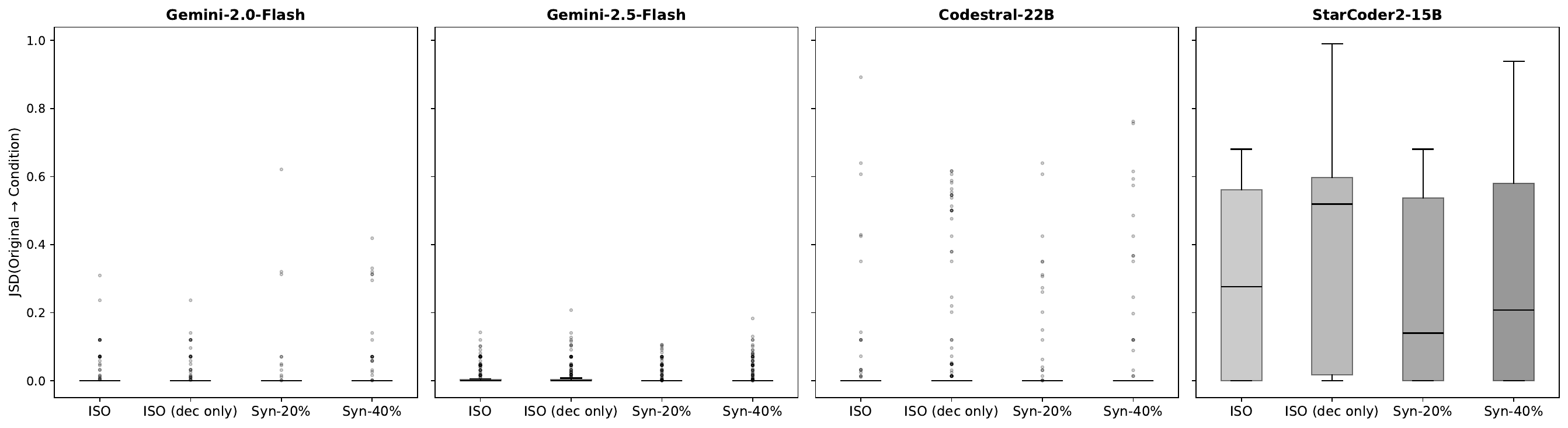}\label{fig:jsd_mbpp}}
\caption{Per-problem Jensen-Shannon divergence of pooled opcode distributions (\textsc{Condition} vs.\ \textsc{Original}).}\label{fig:jsd}
\end{figure}

\subsection{Representational Load Degrades Performance Without Erasing the Algorithm}\label{sec:res_repload}

To answer RQ3 we break it into two smaller sub-questions that can each be tested on their own. A pass@1 drop alone does not pin down the extent to which representational load breaks performance, because the same drop appears both when a model loses the algorithm under the unfamiliar interface and when it keeps the algorithm but mis-serializes the output. RQ3a isolates where in the model the degradation arises, and RQ3b establishes what kind of failure it is.

\subsubsection{A Decoder-Side Bottleneck Drives the Degradation}\label{sec:res_iso_bottleneck}

Table~\ref{tab:passk} reports pass@1 under all evaluation conditions. The full \textsc{Iso} contract drops frontier-model pass@1 by 14--30 absolute points on a provably fixed task, which on its own shows that representational load is sufficient to cause substantial degradation. To localize the source, we decompose the contract into \textsc{Iso (Enc only)}, which transforms only the inputs the model must decode, and \textsc{Iso (Dec only)}, which transforms only the outputs the model must encode. For Gemini-2.0-Flash on MBPP, decoding inputs alone (\textsc{Iso (Enc only)}) costs a modest 9.7 points (80.8$\to$71.1), whereas encoding outputs token by token (\textsc{Iso (Dec only)}) costs a far larger 21.8 points (80.8$\to$59.0), close to the 22.7-point penalty of the full \textsc{Iso} condition (80.8$\to$58.1). This split holds across datasets and across scaled models. The bottleneck is output-side contract compliance rather than input recognition, consistent with autoregressive generation struggling more with formatting constraints held over many decoding steps than the encoder does with parsing a transformed prompt once.

\noindent
\begin{tcolorbox}[colback=gray!5, colframe=black!60]
\textbf{RQ3a.}~\emph{Where does the degradation from representational load originate?}\\[3pt]
\textbf{Primarily on the decoder side.} I/O isomorphisms drop pass@1 by 14--30 absolute points in frontier models, a vulnerability that purely lexical probes such as synonym fuzzing fail to surface. Decomposing the contract shows that most of this penalty comes from output-side encoding rather than input-side decoding, pointing to contract compliance during generation as the dominant cost.
\end{tcolorbox}

\subsubsection{Opcode Analysis Separates Compliance from Competence}\label{sec:res_opcode_analysis}

A pass@1 drop does not say whether the model has lost the algorithm or has kept it and only mis-serialized the interface. We read this distinction off the opcode-entropy analysis, treating a near-zero shift in the opcode distribution as evidence that the underlying strategy is preserved and a large shift as evidence that it is abandoned.

\noindent\textbf{Scaled models preserve core algorithms and control flow across all representational loads.}
For frontier models such as Gemini-2.0-Flash and GPT-4o, the per-problem JSD over all solutions, not just passing ones, under \textsc{Iso} and \textsc{Iso (Dec only)} stays near zero across all three benchmarks (Figure~\ref{fig:jsd}). The bytecode footprint under the metamorphic prompt is close to indistinguishable from that under the original prompt, and the model opts for the same loops, comparisons, and control flow on the exact problems where pass@1 has just dropped by 14--30 points. The combined distributions in Figure~\ref{fig:entropy} track closely between \textsc{Original} and \textsc{Iso}, with the model selecting the same algorithmic families and producing near-identical opcode profiles while fumbling the final output representation more often. As a calibration check, JSD under \textsc{Syn-20} and \textsc{Syn-40} also stays near zero for all models, which indicates that the metric flags preserved core logic rather than any prompt change.

\noindent\textbf{Smaller models lose algorithmic stability as task complexity grows.}
The opcode distributions reveal a contrast between smaller and mid-tier models, with the latter heavily dependent on dataset difficulty. On easier datasets such as MBPP, models like Codestral-22B behave much like scaled models and keep JSD relatively low under \textsc{Iso}. On harder datasets such as BigOBench and EffiBench, however, smaller models buckle under the representational load and their JSD spikes (Figure~\ref{fig:jsd}). This suggests that when the underlying task already sits at the boundary of their capability, the added interface load pushes them to abandon the correct strategy and diverge to qualitatively different and often incorrect solution families, or to fail to produce valid logic at all.

\noindent\textbf{Scale expands algorithmic coverage, not just reliability.}
Comparing the \textsc{Original} opcode profiles \emph{across models} reveals a further dimension of scaling. On MBPP, Gemini-2.0-Flash and Codestral-22B produce overlapping entropy distributions, and the algorithmic repertoire required for basic tasks is within reach of both. On EffiBench and BigOBench the distributions diverge before any perturbation is applied (Figures~\ref{fig:ent_gemini20_eff} and~\ref{fig:ent_codestral_eff}). Gemini-2.0-Flash spans two peaks and performs better at pass@1, while the smaller Codestral-22B largely skips the second peak, consistent with a family of solutions present in the larger model's repertoire but absent from the smaller one.

\noindent
\begin{tcolorbox}[colback=gray!5, colframe=black!60]
\textbf{RQ3b.}~\emph{Is the degradation a competence failure or a compliance failure?}\\[3pt]
\textbf{A compliance failure for scaled models and a competence failure for smaller ones.} Despite large pass@1 drops, frontier models show near-zero opcode JSD under isomorphism, which indicates they attempt the same algorithmic core and mainly fumble the representational channel. Smaller models, by contrast, abandon their strategies under the same load on complex tasks, a pattern more consistent with shallow and fragile task comprehension than with mere mis-serialization.
\end{tcolorbox}

\noindent
\begin{tcolorbox}[colback=gray!5, colframe=black!60]
\textbf{RQ3.}~\emph{To what extent can representational load break performance?}\\[3pt]
\textbf{Sharply, but in scaled models without erasing the algorithm.} The \textsc{Iso} contract drops frontier-model pass@1 by 14--30 absolute points on a provably fixed task, so the loss is the load rather than a memorization artifact. The penalty is concentrated on the decoder, where the model must serialize an encoded output token by token (RQ3a), and for scaled models the algorithmic core survives, surfacing as a narrowed rather than a closed route from specification to code (RQ3b). The competence failure that a bare pass@1 drop might suggest appears only in smaller models on tasks already near their capability limit.
\end{tcolorbox}

\section{Ablation Studies}\label{sec:ablations}

We conduct two ablation studies, one to test whether the observed brittleness generalizes beyond affine transforms, and one to control for prompt-length confounds.

\subsection{Alternative Isomorphism Families}\label{sec:ablation_iso}

\begin{table}[!htbp]
\caption{Pass@1 (\%) under alternative isomorphism families.}\label{tab:ablation_iso}
\scriptsize
\setlength{\tabcolsep}{3pt}
\renewcommand{\arraystretch}{1.15}
\centering
\begin{tabular}{@{}ll rrrr@{}}
\toprule
\textbf{Data} & \textbf{Model} & \textbf{Orig} & \textbf{Affine} & \textbf{Base} & \textbf{Cubic} \\
\midrule
\multirow{3}{*}{BigO}
  & gemini-2.0-flash              & 65.9 & 50.3\,({-}24\%) & 52.2\,({-}21\%) & 52.8\,({-}20\%) \\
  & Codestral-22B-v0.1            & 22.0 & 12.6\,({-}43\%) & 13.4\,({-}39\%) & 11.7\,({-}47\%) \\
  & Qwen2.5-Coder-32B-Instruct    & 30.1 & 18.7\,({-}38\%) & 19.8\,({-}34\%) & 17.5\,({-}42\%) \\
\midrule
\multirow{3}{*}{Effi}
  & gemini-2.0-flash              & 54.4 & 29.3\,({-}46\%) & 26.9\,({-}51\%) & 37.8\,({-}31\%) \\
  & Codestral-22B-v0.1            & 21.3 & 10.7\,({-}50\%) & 12.5\,({-}41\%) &  9.8\,({-}54\%) \\
  & Qwen2.5-Coder-32B-Instruct    & 39.5 & 20.1\,({-}49\%) & 22.5\,({-}43\%) & 18.0\,({-}54\%) \\
\midrule
\multirow{3}{*}{MBPP}
  & gemini-2.0-flash              & 80.8 & 58.1\,({-}28\%) & 52.6\,({-}35\%) & 60.5\,({-}25\%) \\
  & Codestral-22B-v0.1            & 72.0 & 44.1\,({-}39\%) & 49.3\,({-}32\%) & 43.2\,({-}40\%) \\
  & Qwen2.5-Coder-32B-Instruct    & 84.0 & 45.4\,({-}46\%) & 50.5\,({-}40\%) & 43.0\,({-}49\%) \\
\botrule
\end{tabular}
\end{table}

To test whether the brittleness generalizes beyond affine transforms, we ablate with two alternative bijection families targeting distinct representational axes.

\begin{enumerate}
\item \textbf{Base conversion:} The decimal digit string of each integer is treated as if it were written in base~$b$, yielding a different numeric value. For example, with $b{=}7$ the integer $25$ has digit string \texttt{"25"}, which read in base~7 gives $2\!\times\!7+5=19$, so $T(25)=19$. The inverse recovers the original value by inverting the conversion. The base-7 representation of $19$ is \texttt{"25"}, whose digits read in base~10 give $25$. This replaces the arithmetic overhead of affine inversion with a format-parsing overhead, testing whether the model can reason about numeral systems rather than polynomial algebra.
\item \textbf{Cubic polynomials:} $x' = x^3 + c$, requiring explicit cube-root inversion, a qualitatively harder challenge than a linear inverse.
\end{enumerate}

\noindent We run three models across all datasets as shown in Table~\ref{tab:ablation_iso}. All three families produce drops of comparable magnitude. On BigOBench, Gemini-2.0-Flash loses 21\% (base) and 20\% (cubic), closely tracking the 24\% affine drop. On EffiBench, base conversion is steeper than affine ({-}51\% vs.\ {-}46\%), while the cubic transform is milder ({-}31\%). On MBPP, both alternatives again bracket the affine result. The consistency across families confirms that the brittleness reflects general value-space sensitivity rather than an artifact of any particular transform.

\subsection{Prompt Length as a Confound}\label{sec:ablation_length}

\begin{table}[!htbp]
\caption{Pass@1 (\%) under dead-code insertion controlling for the prompt-length confound. Dead-code blocks match the token-count increase of the \textsc{Iso} contract while leaving the task and I/O unchanged.}\label{tab:ablation_length}
\footnotesize
\setlength{\tabcolsep}{3pt}
\renewcommand{\arraystretch}{1.15}
\centering
\begin{tabular}{@{}ll rrr@{}}
\toprule
\textbf{Data} & \textbf{Model} & \textbf{Orig} & \textbf{Iso} & \textbf{Dead Code} \\
\midrule
\multirow{3}{*}{BigO}
  & gemini-2.0-flash            & 65.9 & 50.3\,({-}24\%) & 65.1\,({-}1\%) \\
  & Codestral-22B-v0.1          & 22.0 & 12.6\,({-}43\%) & 20.5\,({-}7\%) \\
  & Llama-3.1-70B-Instruct      & 42.6 & 20.9\,({-}51\%) & 41.0\,({-}4\%) \\
\midrule
\multirow{3}{*}{MBPP}
  & gemini-2.0-flash            & 80.8 & 58.1\,({-}28\%) & 81.4\,(+1\%) \\
  & Codestral-22B-v0.1          & 72.0 & 44.1\,({-}39\%) & 68.5\,({-}5\%) \\
  & Llama-3.1-70B-Instruct      & 71.8 & 41.4\,({-}42\%) & 69.5\,({-}3\%) \\
\botrule
\end{tabular}
\end{table}

The \textsc{Iso} condition appends an encoding contract to the prompt, increasing its length. To rule out prompt length as a confound, we insert semantically irrelevant Python code blocks that match the token-count increase of the \textsc{Iso} contract while leaving the task and I/O unchanged. If length alone drives the observed drops, dead-code insertion should produce comparable degradation. Table~\ref{tab:ablation_length} shows it does not. Dead-code insertion causes negligible drops of 1 to 7\% relative, while \textsc{Iso} degrades the same models by 24 to 51\% on the same benchmarks. The gap is consistent across all model-dataset pairs. On BigOBench, gemini-2.0-flash loses 0.8 points with dead code versus 15.6 points under \textsc{Iso}, and Llama-3.1-70B loses 1.6 points versus 21.7. On MBPP, Gemini-2.0 under dead code actually gains 0.6 points (+1\%), a fluctuation we attribute to noise. In every case, the \textsc{Iso} drop exceeds the dead-code drop by an order of magnitude, confirming that the performance degradation is driven by the representational load of contract compliance, not by prompt length.

\section{Discussion}\label{sec:discussion_conclusion}

\subsection{Why the Decoder-Side Bottleneck May Be Architectural}
Our decomposition experiments indicate that decoding accounts for most of the isomorphism penalty (Section~\ref{sec:res_iso_bottleneck}). A plausible reading of this is the asymmetry between prompt understanding and generation. On the encoder side, the model can integrate the contract and task into a coherent representation before producing any output~\cite{vaswani2017attention}, so it can attend to the whole prompt at once. Decoding, by contrast, is sequential, and each token depends only on the prompt and prior outputs~\cite{radford2019language}. To satisfy the encoding contract, the model has to preserve an implicit arithmetic state across many steps while also maintaining correct Python syntax and contract compliance, and this kind of constrained multi-step generation plausibly stresses working memory in ways that standard next-token training does not explicitly support~\cite{dziri2023faith}. The same asymmetry may also help explain why synonym fuzzing saturates at scale (Section~\ref{sec:res_encoder}), since lexical variation is resolved in the context-rich regime where scaled representations tend to collapse equivalent forms~\cite{ethayarajh2019contextual}, whereas isomorphism shifts the load onto the decoder, which must serialize a novel computation token by token. If this account is correct, the bottleneck is closer to architectural than incidental and might persist in autoregressive models even as scale narrows the gap, though our experiments speak to the size of the effect rather than to its mechanism directly.

\subsection{Representational Load as a Practical Test of Generalization}
The term ``generalization'' is highly ambiguous in the code LLM literature. At one extreme, it might require generating \emph{genuinely novel algorithms}, that is, solutions that do not resemble any training instance in structure or strategy. Our work does not study this aspect of generalization, though some benchmarks, such as ARC-AGI~\cite{chollet2024arc, chollet2025arc}, do. At the other extreme, generalization can simply mean producing correct outputs on unseen inputs, which is trivially satisfied by any model that passes a held-out test set (qualitatively different from its training data). We study a point between these, generalization under a representational load the model could not have seen, and our results suggest that scaled models have internalized algorithmic strategies well enough to transport them across interfaces they have never encountered. This looks closer to what Chollet~\cite{chollet2019measure} terms \emph{developer-aware generalization}, the ability to handle situations not anticipated by the system's designers but lying within the span of its learned priors. We read this as a sign that the field's methods for studying generalization may not have kept pace with the models, and it points toward evaluation frameworks that test whether a model can \emph{apply} what it knows under novel conditions rather than only \emph{reproduce} what it has seen.

\subsection{Scope of the Metamorphic Claim}
The biconditional that gives our protocol its sharpness relies on bijective value transforms over numeric test inputs and outputs. Our findings, therefore, characterize code-LLM behavior specifically on numeric programming tasks, and what they indicate there is fairly narrow, that under representational stress, scaled models appear to retain the algorithmic core and mostly fail at re-serializing the encoded interface. Other classes of representational load, such as novel string-formatting contracts, alternate data layouts, or project-local API conventions, are equally important in practice, but the strict metamorphic-testing guarantee does not directly transfer, because constructing a closed bijection on non-numeric input spaces is itself an open problem. Designing metamorphic-style oracles for those loads while preserving the same provable equivalence on the underlying task is a natural extension we leave for future work.

\subsection{Solution Multiplicity Complicates the Memorization Narrative}
Our opcode analysis indicates that scaled models, particularly on BigOBench and EffiBench, span a broad, multi-modal entropy distribution under \textsc{Original} conditions (Section~\ref{sec:res_opcode_analysis}). We interpret these modes as distinct algorithmic families. A sorting problem might admit $O(n \log n)$ divide and conquer, $O(n^2)$ comparison-based, and $O(n)$ counting-based solutions, each with a characteristic opcode signature. Smaller models, by contrast, tend toward narrow, often unimodal distributions, frequently collapsing to a single solution strategy per problem. If this reading holds, the multiplicity has a bearing on how memorization is interpreted, since methodologies targeting the decoder may be confounded when a model can stochastically switch between several ``correct'' solutions.

\section{Threats to Validity}\label{sec:threats}

\noindent\textbf{Construct validity.}
Static opcode entropy is a proxy for algorithmic strategy and not a direct measurement of it. Two programs with the same opcode histogram can differ in runtime behavior, and two behaviourally identical programs can compile to different bytecode, for instance, a list comprehension against an explicit loop. The Jensen-Shannon divergence we report is invariant to opcode ordering, so it tracks the compositional makeup of a solution rather than its control-flow structure, and a model could, in principle, reorganize its control flow while leaving the histogram unchanged. We mitigate these effects by reporting divergence over both passing and all generations and by reading the metric only as relative movement between matched conditions rather than as an absolute fingerprint. A second construct concern is that representational load is observable only through the transforms we instantiate, so the affine, base-conversion, and cubic families stand in for a property we cannot measure directly.

\noindent\textbf{Internal validity.}
\textsc{Iso} lengthens the prompt, and our dead-code ablation (Section~\ref{sec:ablation_length}) bounds the share of degradation attributable to length alone at roughly 7\%. We decode greedily at $T{=}0.0$ to strip sampling variance, which collapses pass@5 toward pass@1 and may understate the headroom a frontier model would show under sampling. The exact wording of the encode-decode contract is a further free variable, since a differently phrased instruction could prime the decoder more or less effectively, and we hold the phrasing fixed rather than searching over it. For the closed-source models, we query hosted endpoints whose system prompts and checkpoint versions are not disclosed, so a silent provider-side change would surface as a behavioral shift we could not attribute. Finally, the \emph{Seen} labels in the CoDeC probe are probable rather than certain, as noted in Section~\ref{sec:decoder_protocol}, though the asymmetry is shared across every model and does not move the trend.

\noindent\textbf{External validity.}
Our isomorphisms cover three integer-valued bijection families, all of our prompts are Python, and the metamorphic guarantee holds only over numeric input and output. Whether the same narrow-channel behavior appears under string-formatting contracts, structured data layouts, or project-local API conventions is untested, and constructing a closed bijection on those spaces is itself an open problem, as we discuss above. Our model panel is dense throughout and excludes mixture-of-experts architectures, where routing may interact with memorization in ways a dense model does not exhibit. The three benchmarks also favor self-contained algorithmic tasks, so our results speak to function-level synthesis rather than repository-scale or agentic coding. Broader transform families, additional languages, MoE models, and larger task units remain future work.

\section{Conclusion}\label{sec:conclusion}

We do not claim that memorization is absent from scaled code LLMs, only that the diagnostics commonly used to detect it appear to stop discriminating at scale. In our experiments, synonym fuzzing and dead-code insertion on the encoder side, and CoDeC on the decoder side, lose much of their signal on frontier models, so a pass@1 drop or a likelihood shift is, on its own, weak evidence of memorization. Holding the algorithmic task fixed under an I/O isomorphism lets us separate representational load from memorization, and scaled models tend to keep the same solution families while mis-serializing the encoded interface.

The takeaway we would offer is practical. Treat a bare pass@$k$ drop as a starting point rather than a memorization finding, separate representational load before attributing any drop to recall, and report solution-space stability alongside correctness. For most software engineering settings the more useful question is not whether a solution was seen in training but whether a model can carry a known algorithm through an unfamiliar interface. And if memorization is to be studied in the stronger sense the term usually implies, we would suggest approaching it from a different vantage point than the surface and likelihood probes inherited from prior work, which give little to build on at current scales.

\backmatter

\section*{Declarations}

\noindent\textbf{Funding.} This research was funded by the Luxembourg National Research Fund (FNR) in collaboration with the University of Luxembourg. % add your FNR grant/reference number here

\noindent\textbf{Competing interests.} The authors have no competing interests to declare that are relevant to the content of this article.

\noindent\textbf{Data availability.} The dataset generated and analysed in this study is openly available in Zenodo at \url{https://doi.org/10.5281/zenodo.19248561}.

\noindent\textbf{Code availability.} The code is openly available at \url{https://github.com/pkrajput/isomorphic_prompts}.

\noindent\textbf{Author contributions.} Prateek Kumar Rajput framed the methodology and wrote the manuscript. Alberick Euraste Djir\'e and Abdoul Aziz Bonkoungou helped run the experiments. Yewei Song and Iyiola Emmanuel Olatunji contributed to reviewing and framing the manuscript. Jacques Klein and Tegawend\'e F. Bissyand\'e served as project advisors and principal investigators. All authors read and approved the final manuscript.


\end{document}